\documentclass[aps,prb,preprint,superscriptaddress,endfloats,showpacs]{revtex4-1}

\usepackage{dcolumn,amsmath,amssymb}
\usepackage{graphicx, mathtools}
\usepackage{epstopdf}
\usepackage{siunitx, bm, times, color}
\usepackage{csquotes}
\begin{document}
	
	\title{Dynamical effects in Bragg coherent x-ray diffraction imaging on finite crystals}
	
	\author{A.G. Shabalin}
	\affiliation{Deutsches Elektronen-Synchrotron DESY, Notkestr. 85, D-22607 Hamburg, Germany}
	\affiliation{Present address: University of California - San Diego, 92093 La Jolla, USA}
	\author{O.M. Yefanov}
	\affiliation{Deutsches Elektronen-Synchrotron DESY, Notkestr. 85, D-22607 Hamburg, Germany}
	\affiliation{Center for Free-Electron Laser Science, DESY, Notkestr. 85, 22607 Hamburg, Germany}
	\author{V.L. Nosik}
	\affiliation{FSRC "Crystallography and Photonics", Russian Academy of Sciences, Leninskii pr. 59, 119333 Moscow, Russia}
	\affiliation{National Research Nuclear University MEPhI (Moscow Engineering Physics Institute), Kashirskoe shosse 31, 115409 Moscow, Russia}
	\author{V.A. Bushuev}
	\affiliation{M.V. Lomonosov Moscow State University, 119991 GSP-1 Moscow, Russia}
	\author{I.A. Vartanyants}
	\altaffiliation[Corresponding author: ]{ivan.vartaniants@desy.de}
	\affiliation{Deutsches Elektronen-Synchrotron DESY, Notkestr. 85, D-22607 Hamburg, Germany}
	\affiliation{National Research Nuclear University MEPhI (Moscow Engineering Physics Institute), Kashirskoe shosse 31, 115409 Moscow, Russia}
	
	\date{\today}
	
	\begin{abstract}
		We present simulations of Bragg Coherent X-ray Diffractive Imaging (CXDI) data from finite crystals in the frame of the dynamical theory of x-ray diffraction.
		The developed approach is based on numerical solution of modified Takagi-Taupin equations and can be applied for modeling of a broad range of x-ray diffraction experiments with finite three-dimensional crystals of arbitrary shape also in the presence of strain.
		We performed simulations for nanocrystals of a cubic and hemispherical shape of different sizes and provided a detailed analysis of artifacts in the Bragg CXDI reconstructions introduced by the dynamical diffraction.
		A convenient way to treat effects of refraction and absorption supported by analytical derivations is described.
		Our results elucidate limitations for the kinematical approach in the Bragg CXDI and suggest a natural criterion to distinguish between kinematical and dynamical cases in coherent x-ray diffraction on a finite crystal.
	\end{abstract}
	
	
	\maketitle
	
\section{Introduction}
\label{sec:intro}

Since its first demonstration~\cite{robinson2001reconstruction, vartanyants2001partial, williams2003three} Bragg Coherent x-ray Diffractive Imaging (CXDI) has become a powerful technique for analysis of microstructure and strain distribution in submicron crystalline samples~\cite{pfeifer2006three, robinson2009coherent, harder2013coherent, ulvestad2015topological, robinson2016materials, ulvestad2016coherent, dzhigaev2016bragg}.
Recently this approach was extended to imaging ultrafast dynamics in nanocrystals using free-electron lasers~\cite{clark2013ultrafast}.
Nowadays there are several actively exploited experimental approaches based on the Bragg CXDI concept, among those are Bragg ptychography~\cite{godard2011three, zhu2015ptychographic, dzhigaev2017nanowire} and Fourier transform holography~\cite{chamard2010three} (see for review of Bragg CXDI methods Ref.~\cite{vartanyants2015coherent}).
In Bragg CXDI technique a finite crystalline sample is illuminated by intense coherent x-ray beam and an interference pattern in the vicinity of a single or several Bragg reflections is recorded~\cite{Shabalin2016_3Dcolloids}.
An inversion of such data from reciprocal to real space by means of three-dimensional (3D) Fourier transformation provides a high-resolution image of a continuous scattering density distribution in the crystal.
The phase of this complex function represents the projection of a local deformation field on the reciprocal lattice vector~\cite{vartanyants2015coherent}.

For most of such experiments the dimensions of considered specimens are rather small, therefore the approximation of a single scattering event for hard x-rays is typically used.
In the theory of x-ray diffraction by crystals this approach is commonly referred to as the kinematical approximation which is valid while the intensity of the diffracted radiation is small in comparison to the intensity of the incident wave~\cite{als2011elements}.
Kinematical description provides a simple expression which allows to calculate scattered amplitude from a finite size crystal as a Fourier transform of its electron density.
Such simplification is not applicable for larger crystals, with the sizes bigger than the so-called extinction length~\cite{batterman1964dynamical, Pinsker1978dynamical, vartaniants2001XSW, authier2001dynamical}, where effects of cross-coupling between the diffracted and transmitted waves, together with refraction and absorption might become significant and affect Bragg CXDI reconstruction.
These effects can be fully described in the frame of the dynamical theory which, however, does not provide a simple analytical expression for the scattering amplitude from a strained crystal of arbitrary shape.
This theory has been extensively developed already for decades~\cite{batterman1964dynamical, Pinsker1978dynamical, vartaniants2001XSW, authier2001dynamical}, but the influence of the dynamical effects on the results of Bragg CXDI has not yet been fully studied up to now.

The dynamical theory of x-ray diffraction considers the interaction of the wave field with the periodic potential of the crystal lattice taking into account all multiple scattering effects.
In this theory one of the most convenient ways to propagate  the transmitted and diffracted components of the wave field through the weakly strained crystalline media is based on a set of differential equations with corresponding boundary conditions.
This approach developed by Takagi and Taupin~\cite{takagi1962dynamical, taupin1967prevision, takagi1969dynamical} describes a general case of the two-beam dynamical diffraction on a perfect or weakly distorted crystal.
An analytical solution of these equations is nontrivial and can be performed for a few specific cases only, such as a crystal plate finite in one dimension, but infinite in two other dimensions~\cite{afanas1971dynamical, authier2001dynamical, vartaniants2001XSW}.
In a recent work~\cite{gorobtsov2016phase} an analytical solution of the Takagi-Taupin equations was found for the phase of the transmitted beam in a quasi-kinematical approximation.
Methods of numerical integration of the Takagi-Taupin equations for simulations of the wave field distribution in the crystal were developed in Refs.~\cite{epelboin1979boundary, gronkowski1991propagation, mocella2003new}.
Rapidly increasing number of publications on coherent x-ray scattering experiments on finite size crystals in recent years resulted in a growing interest to understand the role of the dynamical scattering effects in these experiments.
For example, Darwin recurrence formalism was applied to study dynamical scattering effects in reciprocal space mapping while scattering on a crystal of rectangular cross-section~\cite{kolosov2005numerical, punegov2014darwin, punegov2016bragg}.
A different approach to solve the Takagi-Taupin equations iteratively via a converging series for a finite size crystal was proposed in Ref.~\cite{yan2014x}.
However, there was still no full analysis of the consequences of the dynamical effects on the reconstruction of the crystal shape and strain field.

In this work we present a general model based on a modification of the Takagi-Taupin equations optimized for geometry of Bragg CXDI measurement aiming to facilitate a numerical solution in a finite 3D crystal of an arbitrary shape in the presence of deformations.
Using this model we performed a series of calculations of 3D maps of the complex scattered amplitude distribution in the surrounding of a Bragg reflection for cube- and hemisphere-shaped crystals of different sizes.
After kinematical (Fourier) inversion of the simulated 3D reciprocal space data sets into real space the results were compared with their original ones thus revealing a character of the artifacts, introduced by the dynamical diffraction.
Next, we analysed effects of refraction and absorption on the reconstructed shape and phase in real space.
By neglecting coupling term for the transmitted and diffracted beams we found an analytical solution of the Takagi-Taupin equations that allows to separate the contributions of refraction and absorption.
We also determined a correction function that allows to eliminate these effects in the Bragg CXDI.
A similar approach was previously presented in Ref.~\cite{harder2007orientation} where the refraction phase shift was calculated accordingly to the optical path for each position in the crystal and subtracted from the results of reconstruction.
In this work we provide detailed analytical consideration and demonstrate our method of correction on simulations.
We also discuss limitations of the kinematical approach in the Bragg CXDI method.

\section{ Bragg CXDI technique}
\label{sec:ConvCXDI}

Typical geometry of a Bragg CXDI experiment assumes an isolated crystal fully illuminated by a coherent x-ray beam.
The size of a crystal is generally significantly smaller than the beam and incoming wave field is considered to be a plane wave~\cite{vartanyants2015coherent}.
The diffracted intensities are recorded by a two-dimensional (2D) pixelated detector located in the far-field and series of diffraction patterns are measured by rotating the sample in the region around the Bragg peak.
In kinematical approximation the complex scattered amplitude $A(\textbf{q})$ in the vicinity of a reflection with the corresponding reciprocal lattice vector $\mathbf h$ is given by a Fourier integral~\cite{vartanyants2015coherent}
\begin{equation}
A(\mathbf{q}) \propto\frac{F_h}{V_{u.c.}} \int S_h(\mathbf r) e^{-i\mathbf{q}\cdot\mathbf r} d\mathbf r \ .
\label{eq:ScattEq}
\end{equation}
Here $F_h$ is the structure factor, $V_{u.c.}$ is the volume of the unit cell, the momentum transfer vector $\mathbf{q}$ is defined as $\mathbf{q}=\mathbf{Q}-\mathbf{h}$, where $\mathbf{Q}=\mathbf{k}_h-\mathbf{k}_0$.
In kinematical approximation both incoming $\mathbf{k}_0$ and diffracted $\mathbf{k}_h$ vectors are defined in vacuum and have the magnitude $|\mathbf{k}_0| = |\mathbf{k}_h| = 2\pi/\lambda$, where $\lambda$ is the wavelength of radiation.
In Eq.~\eqref{eq:ScattEq} we introduced a complex crystalline function
\begin{equation}
S_h(\mathbf r) = s_h(\mathbf r) e^{i\varphi_h(\mathbf r)} \ ,
\varphi_h(\mathbf r) = - \mathbf h \cdot \mathbf u(\mathbf r) \ ,
\label{eq:CrystFunct}
\end{equation}
where its amplitude $s_h(\mathbf r)$ is so-called shape function, that is defined as unity within the crystal and zero everywhere outside it and its phase $\varphi_h(\mathbf r)$ is proportional to the local deformation field $\mathbf u(\mathbf r)$ that describes displacement of atoms from the ideal lattice positions.
In the case of a perfect crystal the intensity distribution function given by a square modulus of the expression~(\ref{eq:ScattEq}) is centrosymmetric with respect to the specific reciprocal lattice nodes.
However, in the presence of a deformation field this symmetry breaks down thus encoding information about the lattice deformations~\cite{vartanyants2015coherent}.

Equation~\eqref{eq:ScattEq} is a basic concept of the Bragg CXDI method.
In particular, it directly shows that the reconstructed complex crystalline function has its amplitude $s_h(\mathbf{r})$ that is determined by the shape function of the crystalline part of the sample and the phase $\varphi_h(\mathbf{r})$.
Here we want to point out that the shape function introduced in Eq.~\eqref{eq:ScattEq} does not give any information about electron density of the sample.
Such information can be deduced only from the CXDI forward scattering experiments (see for review Ref.~\cite{Nugent2010review}).
Variations of the values of the shape function inside the crystal describe rather modulations of atomic planes associated with the chosen reflection and not electron density modulations.
To distinguish between these two cases in the following we will call the shape function also crystalline function.
The phase introduced in Eq.~\eqref{eq:CrystFunct} by its definition $\varphi_h (\mathbf r) =  -\mathbf h\cdot\mathbf u(\mathbf r)$ can be attributed to the projection of the local displacement field on the reciprocal lattice vector $\textbf{h}$ around which the measurements are performed.
The negative sign of the phase reflects the fact that the positive displacement (expanded lattice) leads to the positional shift of the Bragg peak towards lower momentum transfer values of $ \mathbf Q $.
Taking into account that there is an ambiguity in the constant shift of the phase in the phase retrieval, typically it is the difference between the strained and relaxed parts of the crystal that is determined in CXDI experiment and not its absolute value.
It is important to note here that in kinematical approximation described by Eq.~~\eqref{eq:ScattEq} effects of refraction and absorption are not included.

In a typical Bragg CXDI experiment at synchrotron sources 3D measurements of the scattered intensity in the vicinity of the Bragg peak are obtained by an angular scan of the sample with the fixed directions of the incident beam and detector.
The concept of such measurement in reciprocal space is depicted in Fig.~\ref{fig:EwaldSphereScan}.
%
%
If Bragg conditions are exactly satisfied the Ewald sphere crosses the selected reciprocal lattice node.
At this specific angular position of the crystal the momentum transfer vector $ \mathbf Q $ coincides with the reciprocal lattice vector
$ \mathbf h $.
When the crystal is rotated by an angle $ \Delta \theta $, end of the reciprocal lattice vector moves by
$ \Delta \mathbf q =\mathbf h' - \mathbf h$, where $\mathbf h'$ is the reciprocal lattice vector at the new crystal orientation.
Typical values of the angular deviation in Bragg CXDI experiments do not exceed one degree, therefore the length of the vector
$\Delta \mathbf q$ can be well approximated as $ |\mathbf h| \Delta \theta $.
In our formalism we assume that directions and magnitudes of the incident and diffracted wave vectors $ \mathbf k_0 $, $ \mathbf k_h $ are constant during the rocking scan and they always form a constant angle $2 \theta_B $ at all values of the angular deviation
$ \Delta \theta $.
As such, the wave vector of diffracted field $ \mathbf k_h $ is defined as a constant vector of magnitude $2\pi/\lambda$ pointing at that position on the Ewald sphere which crosses the reciprocal lattice node when the Bragg condition is exactly fulfilled
\begin{equation}
\label{eq:kd_def}
\mathbf k_h=\mathbf k_0+\mathbf h|_{\Delta \theta =0} \ .
\end{equation}

\section{ Model description}
\label{sec:ConvCXDI}

Conventional Bragg CXDI is based on Eq.~\eqref{eq:ScattEq} which is valid only in the frame of kinematical approximation.
For large crystals the kinematical description breaks down and equation~\eqref{eq:ScattEq} cannot be used any more.
In this section we will discuss how this simple approach can be modified when dynamical scattering effects are taken into account.

Here dynamical simulations of the Bragg CXDI will be performed in the geometry described in the previous section (see Fig.~\ref{fig:EwaldSphereScan}).
A detailed sketch of the implemented numerical model is presented in Fig.~\ref{fig:Scheme_NumInt}.
As shown in this figure, the crystal is embedded in a 3D shape rhombic prism formed by the directions of the incident and diffracted vectors $\mathbf k_0$ and $\mathbf k_h$ and centered around the crystal rotation axis.
Simulations of the scattered amplitudes for each value of the angular deviation $\Delta \theta$ are performed in four steps.
On the first step, the 2D distribution of the incoming wave field $E_{in}(\mathbf{r})$ is projected on the left facet of the rhombic prism (shown as $s_0^{in}$ in Fig.~\ref{fig:Scheme_NumInt}).
Next, the wavefield is propagated through the scattering volume by numerical solution of the Takagi-Taupin equations~\cite{takagi1962dynamical,taupin1967prevision,takagi1969dynamical}.
As a result corresponding 2D distribution of the transmitted $E_0(\mathbf{r})$ and diffracted $E_h(\mathbf{r})$ amplitudes is obtained at the exit facets of the rhomb (at $s_0^{out}$ and $s_h^{out}$ facets in Fig.~\ref{fig:Scheme_NumInt}).
At the third step, thus determined distribution of the diffracted amplitude $E_h(\mathbf{r})$ is projected on the plane, perpendicular to the diffracted wave vector $\mathbf k_h$, yielding the 2D exit surface wave (ESW) $E_h^{ESW}(\mathbf{r})$.
Finally, on the fourth step, the ESW is propagated to the detector plane, which in the far-field (Fraunhofer) limit is obtained by applying 2D Fourier transform~\cite{Note1}.
When all series of 2D diffraction patterns as a function of the rocking angle $\Delta\theta$ are simulated, we merge them all together into a 3D scattered intensity map in reciprocal space.
By this simulation of the intensity distribution in the far-field including dynamical effects in scattering are finalised.

In order to understand what kind of artifacts are introduced by the dynamical scattering the simulated intensity distribution have to be inverted to real space by applying the phase retrieval techniques~\cite{Fienup1982phase, Marchesini2007phase}.
In our case, as soon as the complex amplitudes in the far-field are known, they can be directly inverted to real space without the need of the phase retrieval procedure.
Characterization of the dynamical artifacts in real space is performed by a comparison of the output of these simulations with the original crystalline function $S_h(\mathbf{r})$ (see Eq.~\eqref{eq:CrystFunct}).
In the next sections we describe all these steps in more details.

\subsection{Propagation of the wave field through the crystal}

In our simulations of Bragg CXDI we used the laboratory coordinate system in which the direction of the incident beam and detector position are fixed during the angular scan and the sample is rotating.
The origin of the coordinate system is chosen on the crystal rotation axis that is parallel to the y-axis (see Fig.~\ref{fig:Scheme_NumInt}).
In the following we assume the two-beam diffraction conditions
\begin{equation}
\mathbf{E}(\mathbf r) = \sum_s \left[ \mathbf{e}_{0s} E_{0s}(\mathbf r)e^{i\mathbf k_0\cdot\mathbf r}
                    + \mathbf{e}_{hs} E_{hs}(\mathbf r)e^{i\mathbf k_h\cdot\mathbf r} \right] \ ,
\label{eq:TwoWaveApp}
\end{equation}
where $\mathbf{e}_{0s}$ and $\mathbf{e}_{hs}$ are the polarization unit vectors and $s$ is the polarization index.
As a sample we consider a perfect or weakly deformed finite size crystal.
To propagate the complex electric field through the volume of the three-dimensional crystal we introduce the symmetric form of the Takagi-Taupin equations (see Appendix~\ref{app:AppT} for details)
\begin{equation}
\begin{split}
\frac{\partial E_{0s}(\mathbf r)}{\partial s_0}  =  \frac{i\pi}{\lambda} [\chi_0 E_{0s}(\mathbf r)+C\chi_{\bar h} e^{-i\Delta \mathbf q\cdot \mathbf r+i\mathbf h\cdot\mathbf u(\mathbf r)} E_{hs}(\mathbf r)] \ , \\
\frac{\partial E_{hs}(\mathbf r)}{\partial s_h}  =  \frac{i\pi}{\lambda} [\chi_0 E_{hs}(\mathbf r)+C\chi_h e^{i\Delta \mathbf q\cdot \mathbf r-i\mathbf h\cdot\mathbf u(\mathbf r)} E_{0s}(\mathbf r)] \ ,
\end{split}
\label{eq:TTMod}
\end{equation}
%
which are supplemented by the boundary conditions.
Here the partial derivatives $\partial/\partial s_0, \partial/\partial s_h$ are taken along the directions of the wave vectors $\mathbf k_0$ and $\mathbf k_h$.
Response of the crystal is described by the Fourier components of the susceptibility $\chi_0=\chi_{0r}+i\chi_{0i}$ and $\chi_{h, \bar h} = \chi_{hr, \bar hr}+i\chi_{hi, \bar hi}$.
Real and imaginary parts of the zeroth component of the susceptibility $\chi_{0r}$,$\chi_{0i}$ describe effects of refraction and absorption, respectively.
The term with $ \chi_h $ describes diffraction of the transmitted component $E_{0s}(\mathbf r)$ by a set of crystallographic planes with the reciprocal vector $ \textbf{h} $.
In its turn, the diffracted component $E_{hs}(\mathbf r)$ undergoes diffraction by the same set of planes but from the opposite side, which is described by the term with $ \chi_{\bar h} $.
Vector $ \Delta \mathbf q$ determines the angular deviation from the exact Bragg condition as shown in Fig.~\ref{fig:EwaldSphereScan}.
%
%
In equations~(\ref{eq:TTMod}) $ C $ stands for the polarization factor, which is equal to unity in the case of a $\sigma-$polarization and $\cos2\theta_B$ in the case of a $\pi-$polarization.
Without restricting the generality in the following we will assume only $\sigma-$polarization with $C=1$ and omit polarization index $s$ in the wavefield amplitudes.

The boundary conditions assume that the total electric field, represented by equation~(\ref{eq:TwoWaveApp}) is continuous everywhere on the crystal-vacuum boundary.
In our formalism, similar to Refs~\cite{takagi1962dynamical, takagi1969dynamical, taupin1967prevision}, we assume that the wave vectors $\mathbf k_{0,h}$ are the same inside and outside the crystal, therefore the amplitudes
$E_{0,h}(\mathbf r)$ are continuous functions on the boundary of the crystal.
Consequently, equations~(\ref{eq:TTMod}) do not require any specific transformation of the amplitudes $E_{0,h}(\mathbf r)$ on the crystal-vacuum boundary which is particularly convenient in the case of a three-dimensional crystal with an arbitrary shape.
We note that the Fourier components  $\chi_{0,h, \bar h}$ of susceptibility drop down to zero outside the crystal and thereby undergo discontinuity on the crystal boundary.

For a finite size crystal the evolution of the wave field depends on the crystals size, shape, and diffraction geometry.
When exact Bragg conditions are satisfied, the transfer of energy from the transmitted beam into the diffracted beam is strongly enhanced due to constructive interference of the wavefield inside the crystal.
At the same time at these conditions the wave field is not penetrating deep into crystal.
This effect, known as extinction~\cite{Pinsker1978dynamical, authier2001dynamical}, is described by the characteristic decay length of the wave field commonly referred to as the extinction length $ L_{ex}$~\cite{Vartaniants2001,Note2}
\begin{equation}
L_{ex} = \frac{\lambda\sqrt{\gamma_0|\gamma_h|}}{\pi \mathit{Re}\left[\sqrt{\chi_{h}\chi_{\bar{h}}}\right]} \ ,
\label{eq:ExtLength}
\end{equation}
where $\gamma_{0,h} = \cos(\mathbf n \cdot \mathbf k_{0,h})$ are the direction cosines and $\mathbf n$ is the inward normal to the entrance surface of the crystal.
In the denominator of the expression~\eqref{eq:ExtLength} the real part of the complex valued square root $\sqrt{\chi_{h}\chi_{\bar{h}}}$ is used.
The extinction length (\ref{eq:ExtLength}) is commonly referred to as a characteristic value to distinguish between the cases of the kinematical and dynamical diffraction.
When the crystal size is much smaller than the extinction length the effects of coupling between the transmitted and diffracted components of the wave field are small and kinematical approximation can be used safely.
When the crystal size is about or bigger than the extinction length these effects are becoming important and the dynamical theory has to be used.

\subsection{Numerical solution of the Takagi-Taupin equations}

In this work we perform numerical integration of the Takagi-Taupin equations~\eqref{eq:TTMod} applying an approach similar to that described in Ref.~\cite{gronkowski1991propagation}.
%
%
To propagate the wavefield along the directions of partial derivatives $\partial/\partial s_0, \partial/\partial s_h$, we introduce the laboratory coordinate system with the origin on the crystal rotation axis.
The set of basis vectors $\{\mathbf s_0, \mathbf s_h, \mathbf s_y \}$ is represented by the unit vectors in the direction of the incident beam ($\mathbf s_0$), diffracted beam ($\mathbf s_h$) and normal to the scattering plane ($\mathbf s_y$) (see Fig.~\ref{fig:Scheme_NumInt}).
Thus, the partial derivatives are taken along $\mathbf s_0$ and $\mathbf s_h$ vectors, and rotation is performed around the $\mathbf s_y$-axis.
The angle between the vectors $\mathbf s_0$ and $\mathbf s_h$ is equal to $\num{2}\theta_B$, therefore the coordinate system generally is not orthogonal.
Any position within the considered volume can be described by the radius vector $\mathbf r = s_0\mathbf s_0 + s_h\mathbf s_h + s_y\mathbf s_y$, where $s_{0,h,y}$ are corresponding coordinates.

We perform the numerical integration over a rhombic prism in which the whole crystal is embedded, as shown in Fig.~\ref{fig:Scheme_NumInt}.
More specifically, the prism is sliced to a set of layers, defined for different values of $s_y$ coordinate parallel to the scattering plane and Takagi-Taupin equations~\eqref{eq:TTMod} are solved in the two-dimensional grid independently for each of these layers.
Since directions $\mathbf s_{0,h}$ do not depend on the angular deviation $\Delta\theta$ in chosen coordinate system the whole grid remains invariable during the angular scan, while rotation transformations are applied to the susceptibility and shape function of the crystal (see Appendix~\ref{app:AppT} for details).
The nodes, which belong to the crystal, are characterized by the values of Fourier components of the susceptibility, which are replaced by zeros for the nodes outside the crystal.

In the numerical integration method the complex amplitudes $E_{0,h}(\mathbf r)$ are represented by a discrete set of values over all integration grid and the Takagi-Taupin equations~\eqref{eq:TTMod} are transformed into a pair of recurrence relations (see Appendix~\ref{app:num} for details).
The inset in Fig.~\ref{fig:Scheme_NumInt} shows the recurrence property of the obtained equations for the neighboring nodes of the integration grid.
For the node $(i,j)$ the values of amplitudes $E_{0, h}^{(i,j)}$ are calculated from the values $E_{0, h}^{(i-1,j)}$ and $E_{0, h}^{(i,j-1)}$ at the previous nodes $(i-1,j)$ and  $(i,j-1)$.
In such a way calculations proceed from node to node in the direction of the transmitted and diffracted beams.
The values of the amplitudes $E_{0,h}(\mathbf r)$ on the left ($s_0^{in}$) and bottom ($s_h^{in}$) sides of the prism are defined as
\begin{equation}
E_0(\mathbf r) = E_{in}(\mathbf{r}), \ \text{at} \ s_0=s_0^{in} \ \text{and}
\ E_h(\mathbf r) = 0,  \ \text{at} \ s_h=s_h^{in} \ .
\label{eq:BCond}
\end{equation}
Such form of boundary conditions is universal and particularly convenient to implement for numerical integration in the case of a three-dimensional crystal with an arbitrary shape.
Once established these boundary conditions can be applied to any shape and orientation of the crystal embedded into the integration prism.

\subsection{Propagation to the detector plane}

Numerical integration of the Takagi-Taupin equations~(\ref{eq:TTMod}) over the rhombic prism results in the complex amplitude of the transmitted beam $E_0(\mathbf{r})$ at the right facet of the prism ($s_0^{out}$) and
diffracted wave $E_h(\mathbf{r})$ at the upper facet of the prism ($s_h^{out}$).
For the free space propagation to the detector we exploit an orthogonal coordinate system with the basis $\{\mathbf s_{\bot}, \mathbf s_h, \mathbf s_y\}$, where the vector $\mathbf s_{\bot}=\mathbf s_h \times \mathbf s_y$ is introduced (see Fig.~\ref{fig:Scheme_NumInt}).
This vector is perpendicular to the direction of propagation for diffracted component and lies in the scattering plane, therefore the transition for the coordinate $s_0$ is performed by means of a simple projection $s_{\bot}=s_0\sin(\num{2}\theta_B)$.
The result of such projection, applied to the calculated 2D distribution of the diffracted wave field will be further referred to as the exit surface wave $E_h^{ESW}(s_{\bot},s_y, \Delta \mathbf q)$.
%
%
To determine the scattered amplitude at the detector plane in the far-field we apply 2D Fourier transform to the exit surface wave
\begin{equation}
\begin{split}
A(q_{\bot},q_y, \Delta \mathbf q) =
\iint E_h^{ESW}(s_{\bot},s_y, \Delta \mathbf q) e^{-iq_{\bot}s_{\bot}-iq_y s_y}ds_{\bot}ds_y \ ,
\end{split}
\label{eq:FT_EW}
\end{equation}
where $q_{\bot}, q_y$ are corresponding coordinates in reciprocal space.
%



\begin{figure}[p]
	\centerline{\includegraphics[angle=0, width=8.6cm, trim={1.5cm 11.0cm 1.5cm 1.5cm}, clip]{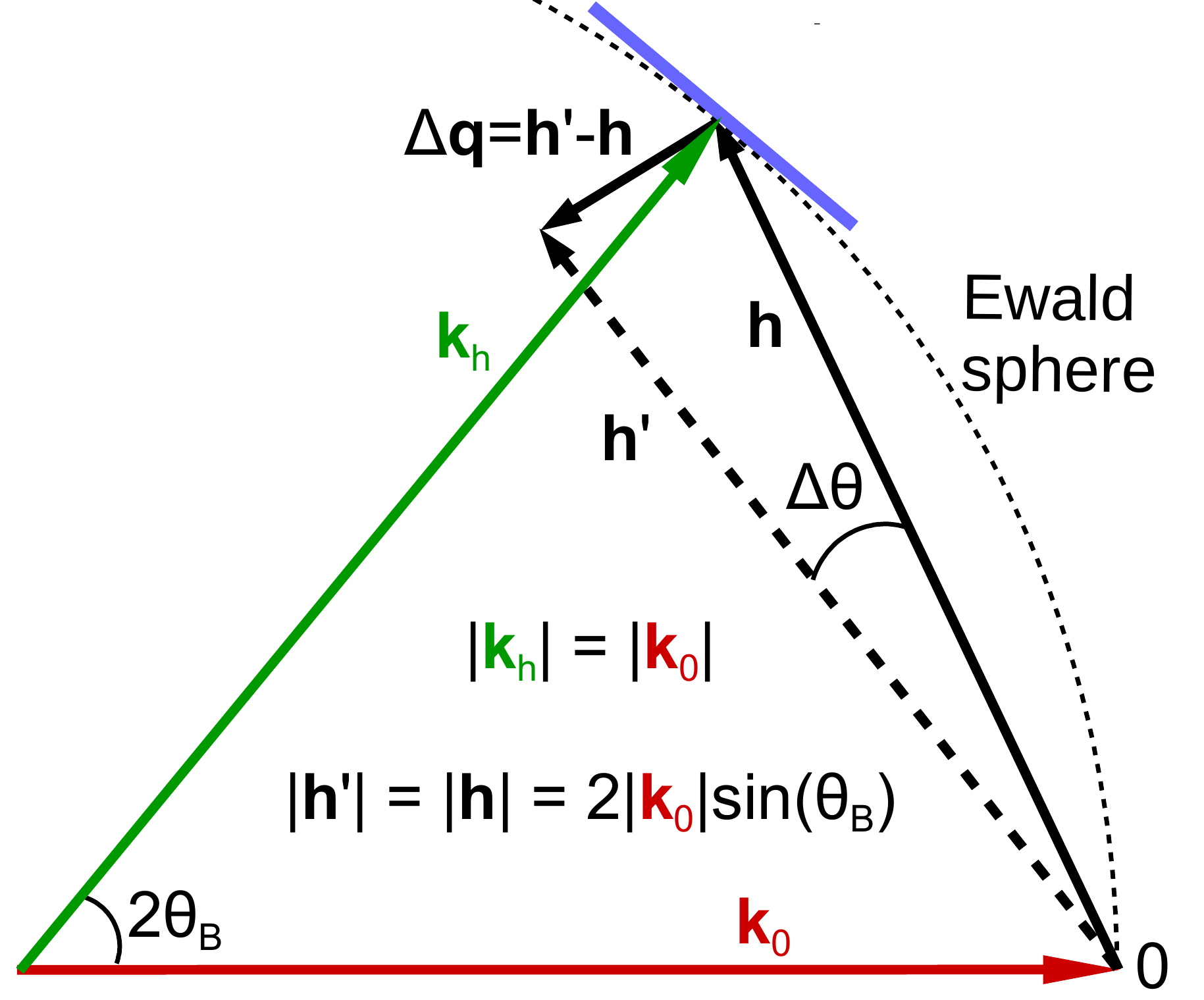}}
	\caption{
		Geometry of the Bragg CXDI measurement.
		The wave vector of the diffracted beam $ \mathbf k_h = \mathbf k_0+\mathbf h$ is composed by the wave vector of the incident beam
$ \mathbf k_0 $ and reciprocal lattice vector $ \mathbf h $ at a specific crystal orientation when exact Bragg conditions are satisfied.
		The vectors $ \mathbf k_h $ and $ \mathbf k_0 $ form an angle of $2\theta_B$.
		The vector $ \mathbf h' $ corresponds to the crystal reciprocal lattice vector while the crystal is rotated by an angle $ \Delta \theta$.
		The diffraction pattern, recorded by the 2D detector, maps a part of the spherical surface in reciprocal space described by the Ewald sphere.
        Blue line indicates approximation of the Ewald sphere by flat surface.
        }
    	\label{fig:EwaldSphereScan}
\newpage
\end{figure}

In generic Bragg CXDI experiment the measured diffraction pattern corresponds to the cut of reciprocal space by the Ewald sphere (see Fig.~\ref{fig:EwaldSphereScan}).
Our model does not account for divergence of the wave field $E_{0,h}(\mathbf r)$ while its propagation in a crystal (see Appendix~\ref{app:AppT}).
In fact, this is similar to the projection approximation, when the simulated 2D diffraction pattern is attributed to the flat surface in reciprocal space (see Fig.~\ref{fig:EwaldSphereScan}).
As such, the 2D distribution of the scattered amplitude $A(q_{\bot},q_y, \Delta \mathbf q)$ defined by Eq.~\eqref{eq:FT_EW} is determined in a plane in reciprocal space perpendicular to the direction of the diffracted wave $\mathbf{s_h}$ and corresponding to a fixed angular deviation $\Delta \mathbf q$.
Changing the value of the rocking angle $\Delta\theta$ the full set of complex amplitudes $A(q_{\bot},q_y, \Delta \mathbf q)$ is determined in reciprocal space.
By taking the square modulus of the amplitudes the final 3D distribution of the intensity
$I(q_{\bot},q_y, \Delta \mathbf q) = \left|A(q_{\bot},q_y, \Delta \mathbf q)\right|^2$ is obtained.
As a next step, this set of 2D images is interpolated on a 3D uniform grid with the orthogonal coordinates $q_x, q_y, q_z$ (see Fig.~\ref{fig:ReciAndRealSketch}(a)).

As soon as phases of the scattered amplitudes are known in our simulations we perform 3D inverse Fourier transform of simulated 3D amplitudes
\begin{equation}
S_h(\mathbf{r}) =  s_h(\mathbf r) e^{i\varphi_h(\mathbf r)} = \frac{1}{{(2\pi)}^3}\int A(\mathbf{q}) e^{i\mathbf{q} \cdot \mathbf{r}} d\mathbf{q} \ ,
\label{eq:Inversion}
\end{equation}
where $S_h(\mathbf{r})$ is a complex crystalline function defined in \eqref{eq:CrystFunct}.
In kinematical approximation it should reproduce the crystal shape by its amplitude and be proportional to the projected strain field by its phase (compare with Eq.~\eqref{eq:CrystFunct}).
According to this approach, if a crystal is unstrained, the inversion of the scattered amplitude by Eq. \eqref{eq:Inversion} should give a real shape function with the constant amplitude values.
We will see in the following how dynamical scattering may affect these results.

\section{ Results}
\label{sec:results}

\begin{figure}
	\centerline{\includegraphics[angle=0, width=8.6cm, trim={1.2cm 17.7cm 1.2cm 1.0cm}, clip]{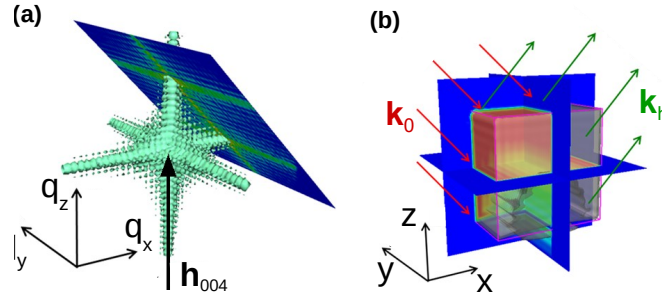}}
	\caption{
		Diffraction geometry considered in simulations for a Au crystal of a cubic shape.
        (a) Scattered amplitude distribution in reciprocal space (logarithmic scale) calculated by numerical integration of the Takagi-Taupin equations.
		The tilted plane illustrates amplitude distribution within one of the diffraction patterns at the fixed angular deviation value $\Delta\theta$.
		(b) Schematic view of the diffraction geometry in real space.
        Red arrows indicate the incident beam and green arrows the diffracted beam.
        Results of inversion from reciprocal to real space are also shown here by different colors in transverse slices (see text for details).
	}
	\label{fig:ReciAndRealSketch}
\newpage
\end{figure}

In order to illustrate general features of the dynamical scattering effects in the Bragg CXDI we considered  first a simple object in the form of a cubic-shaped gold crystal without strain.
A schematics of the diffraction geometry in real and reciprocal space and the orthogonal coordinate system with the $x,y,z$-axes oriented along the cube edges is shown in Fig.~\ref{fig:ReciAndRealSketch}.
We assume that a cubic unit cell (with a lattice parameter $a=$\SI{4.078}{\angstrom}) is also aligned along the same coordinate axes.
In our simulations we considered the incident plane wave with \SI{8}{\kilo\electronvolt} photon energy (wavelength $\lambda=$\SI{1.55}{\angstrom}) and 004 reflection conditions.
In this scattering geometry reciprocal space vector $\mathbf{h}_{004}$ is parallel to $q_z-$axis in reciprocal space (see Fig.~\ref{fig:ReciAndRealSketch}(a)) and scattering plane is parallel to $xz-$plane in real space (see Fig.~\ref{fig:ReciAndRealSketch}(b)).
The Bragg angle in these conditions is $\theta_B = 49.47^{\circ}$ and values of the extinction length $L_{ex}$~(\ref{eq:ExtLength}) are
$711$ nm and $607$ nm in the Bragg and Laue geometry, respectively.

\begin{figure}
	\centerline{\includegraphics[angle=0, width=8.6cm, trim={1.5cm 0cm 1.0cm 1.5cm}, clip]{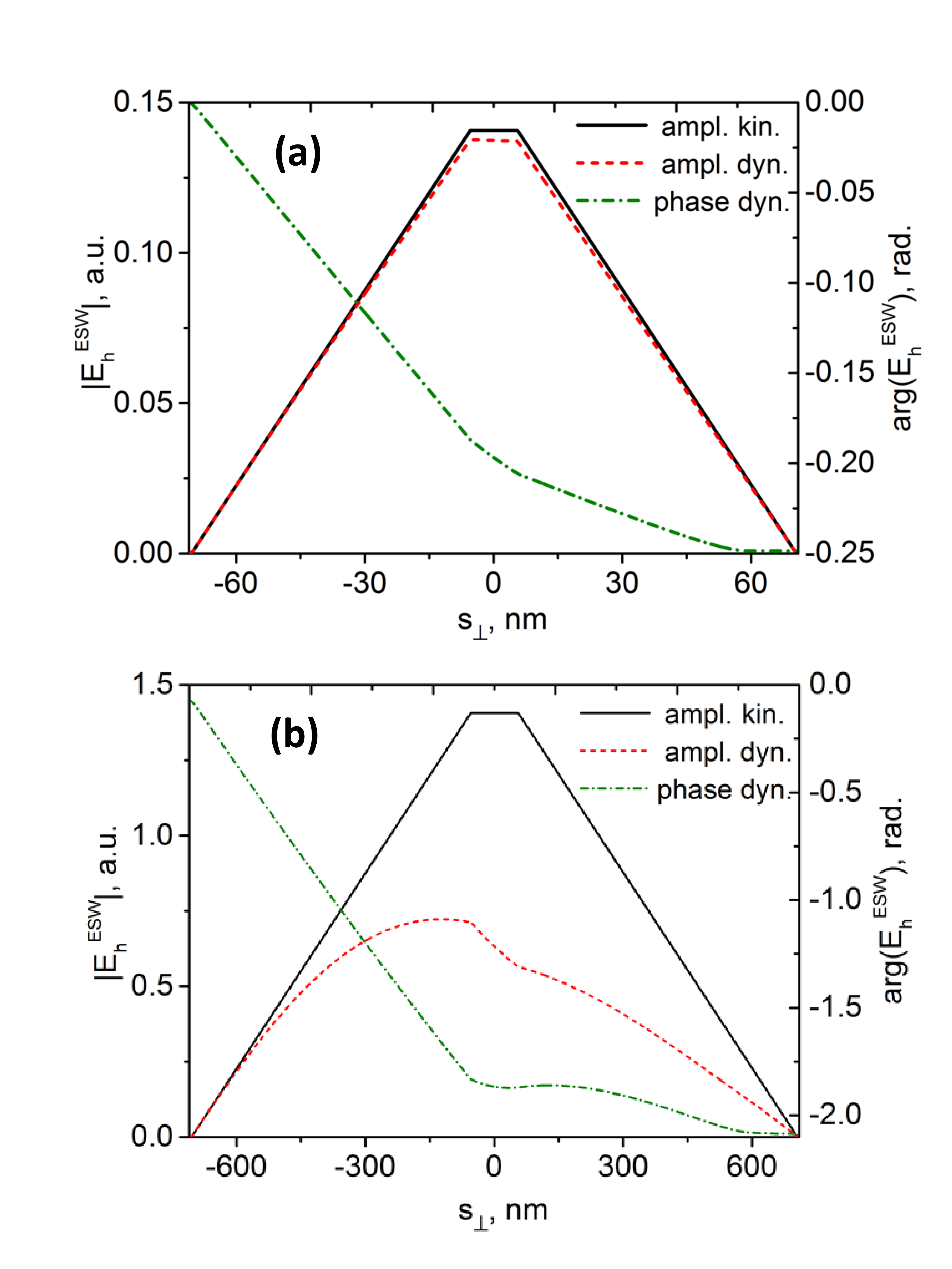}}
	\caption{
		(a) Transverse profile of the amplitude (red) and phase (green) of the exit surface wave $E_h^{ESW}(s_{\bot},\Delta \mathbf{q} =0)$ calculated by the dynamical theory for a \SI{100}{\nano\meter} cubic crystal of Au at exact Bragg conditions.
		For comparison, the amplitude profile obtained in the frame of the kinematical theory is shown by the black curve.
		(b) Same for a crystal of \SI{1}{\micro\meter} size.
	}
		\label{fig:ExitWave_AuCube}
\newpage
\end{figure}

We performed simulations for two crystal sizes of \SI{100}{\nano\meter} and \SI{1}{\micro\meter}.
For \SI{100}{\nano\meter} crystal the angular scan was performed covering the angular range from $-3.3^\circ$ to $+3.3^\circ$, with the angular increment of $6\cdot 10^{-3}$ degree.
The exit surface wave field distribution $E_h^{ESW}(s_{\bot}, \Delta \mathbf{q}=0)$ obtained by a solution of Takagi-Taupin equations at exact Bragg conditions is presented in Fig.~\ref{fig:ExitWave_AuCube}.
The amplitude of the exit wave calculated in the frame of the dynamical theory (red curve) is compared to the results of the kinematical theory (black curve) obtained by setting $\chi_{0,\bar{h}}=0$ (see Appendix\ref{app:AnSol}).
Nearly complete coincidence in simulations for a \SI{100}{\nano\meter} crystal (see Fig.~\ref{fig:ExitWave_AuCube}(a)) suggests that the cross coupling between the diffracted and transmitted waves is not strong enough to have any significant effect on the scattering and, therefore, kinematical approximation provides a rather accurate result.
The calculated phase profile (green curve) shows a small phase shift, which can be attributed to refraction.
The phase distribution is shown relative to the phase of the incoming wave that was set to zero, so that the phase of the diffracted wave at the top left corner of the crystal (see Fig.~\ref{fig:ReciAndRealSketch}) appears to be zero as well.
As a result of our simulations we can see that the phase due to refraction accumulates more for the waves propagating the longest distance in the crystal from its depth and finally reaches its minimum value of \SI{-0.25}{\radian}.

Similar simulations performed for a \SI{1}{\micro\meter} Au crystal are presented in Fig.~\ref{fig:ExitWave_AuCube}(b).
For a \SI{1}{\micro\meter} crystal the angular scan was performed covering the angular range from $-0.83^{\circ}$ to $+0.83^{\circ}$, with the angular increment of $3.3\cdot10^{-3}$ degree.
The dynamical calculations revealed a considerably lower amplitude profile in comparison to the kinematical prediction, which can be attributed to the attenuation of the transmitted wave due to extinction.
This affects mostly the lower part of the crystal.
We would like to note that contribution to attenuation due to normal absorption is much lower than extinction effect.
Indeed, taking into account that normal absorption length for gold at the considered photon energy is \SI{2.9}{\micro\meter}, we obtain attenuation of the x-ray amplitude only by 16\% on the length of a Au particle of \SI{1}{\micro\meter} in size.
The phase distribution, in fact, reproduces major features of the phase for \SI{100}{\nano\meter} crystal, which supports our observation that characteristic phase gradient originate mostly from refraction.
At the same time, the phase profile at the lower part of the cube reveals slight but noticeable bending, which cannot be attributed to refraction, since the refraction phase is linear.
We also notice that due to a bigger crystal size the observed phase shift is about one order of magnitude larger than in the case of \SI{100}{\nano\meter} crystal
and reaches the value of $-2.17$ rad.
Below we will analyse results of inversion obtained for the two different crystal sizes separately.

\subsection{100 nm Au crystal of a cubic shape}

\begin{figure}
	\centerline{\includegraphics[angle=0, width=8.6cm, trim={0.7cm 9.0cm 1.0cm 1.0cm}, clip]{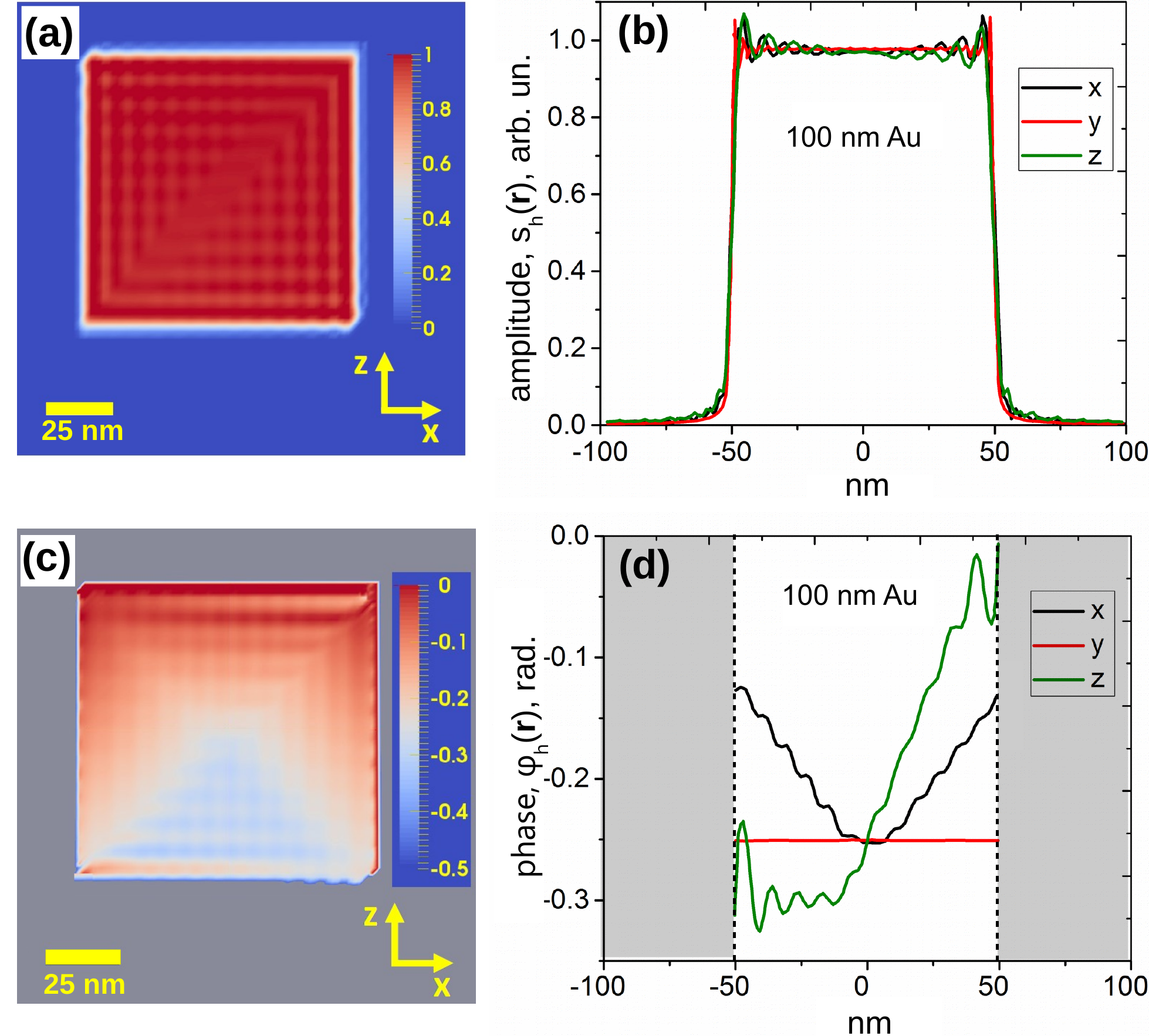}}
	\caption{
		Amplitude (a, b) and phase (c, d) of a complex crystalline function $S_h(\mathbf{r})$ obtained by inversion of the 3D reciprocal space dataset calculated for a crystal size of \SI{100}{\nano\meter}.
		(a,c) The $xz$-slices at $y=\num{0}$ (center of the crystal).
		(b,d) The line profiles through the center of the crystal and along the $x$, $y$, and $z$-axes.
		Gray area in (c, d) outlines region outside the crystal, where the phase is undefined.
	}
	\label{fig:Cube100nm}
\newpage
\end{figure}

Results of inversion of the whole 3D reciprocal space dataset for a \SI{100}{\nano\meter} Au crystal obtained by the dynamical simulations are presented in Fig.~\ref{fig:Cube100nm}.
A 2D distribution of the crystalline amplitude function $s_h(\mathbf{r})$ in $xz$-slice taken through the center ($y $=0) of the crystal is shown in Fig.~\ref{fig:Cube100nm}(a); the line profiles along $ x,y,z$-axes are given in Fig.~\ref{fig:Cube100nm}(b).
The phase distribution $\varphi_h(\mathbf{r})$ is presented in a similar way by the corresponding slice in Fig.~\ref{fig:Cube100nm}(c) and line profiles in Fig.~\ref{fig:Cube100nm}(d), respectively.
Outside of the cube the amplitude of the reconstructed complex density function is rapidly going down (see Fig.~\ref{fig:Cube100nm}(a,b)).
In this region the phases are not defined, therefore, the phase distribution presented in Fig.~\ref{fig:Cube100nm}(c,d) was cropped by the cube edges.

Distribution of the crystalline function reveals, as expected, well defined cubic structure of our model sample.
We should note here that due to plane facets of a cubic sample crystal truncation rods~\cite{als2011elements} are extending quite far in reciprocal space and induce observed oscillations in the crystalline amplitude function obtained by Fourier inversion.
Therefore, slight periodic variations of its values are due to truncation of reciprocal space intensities imposed by the limited range where simulations were performed.

The most intriguing and not expected result was obtained for the phase $\varphi_h(\mathbf{r})$ of the crystalline function (see Fig.~\ref{fig:Cube100nm}(b)).
Instead of being a uniform function inside of an unstrained crystal it shows slight variation of the phase going down to the values of about $-0.3$ rad.
When attributed to strain, these values of the phase would give rise to the displacement of about $0.049$ {\AA} and associated strain for a Au crystalline sample of $~1.2 \cdot 10^{-2}$.
As we will show in the following these variations of the phase can be attributed to refraction effects, that are not considered in the conventional kinematical theory.
Indeed, on the top facet neither incident nor scattered wave experience refraction, therefore the phase shift is zero.
When radiation penetrates deep in the crystal the phase shift due to refraction is accumulated on its way in and out of the crystal.
Since refraction index for x-rays is less then one the accumulated phase is negative.
For the lower facet of the cube the phase shift reaches its minimum value of about \num{-0.3} rad.
This value corresponds to an optical path length of x-ray beam going in and out of the crystal, which for the Bragg angles smaller than $63.43^{\circ}$ gives for the phase shift due to refraction  $\varphi_{refr} = -(2\pi/\lambda)\delta ( d/\cos\theta_B) \simeq -0.29$ rad, where $\delta$ is the real part of the refraction index.
We will introduce later a correction function that will compensate these effects completely and will allow to determine correct values of the phase that can be attributed to strain.

From these simulations we can see that even in the case of very small crystalline samples when dynamical effects should not play any role refraction effects introduce certain phase variations in the reconstructed crystalline function that could lead to a wrong statements about the strain field in the sample.

%

\subsection{1 $\mu$m Au crystal of a cubic shape}

\begin{figure}
	\centerline{\includegraphics[angle=0, width=8.6cm, trim={1.0cm 13.2cm 1.2cm 1.0cm}, clip]{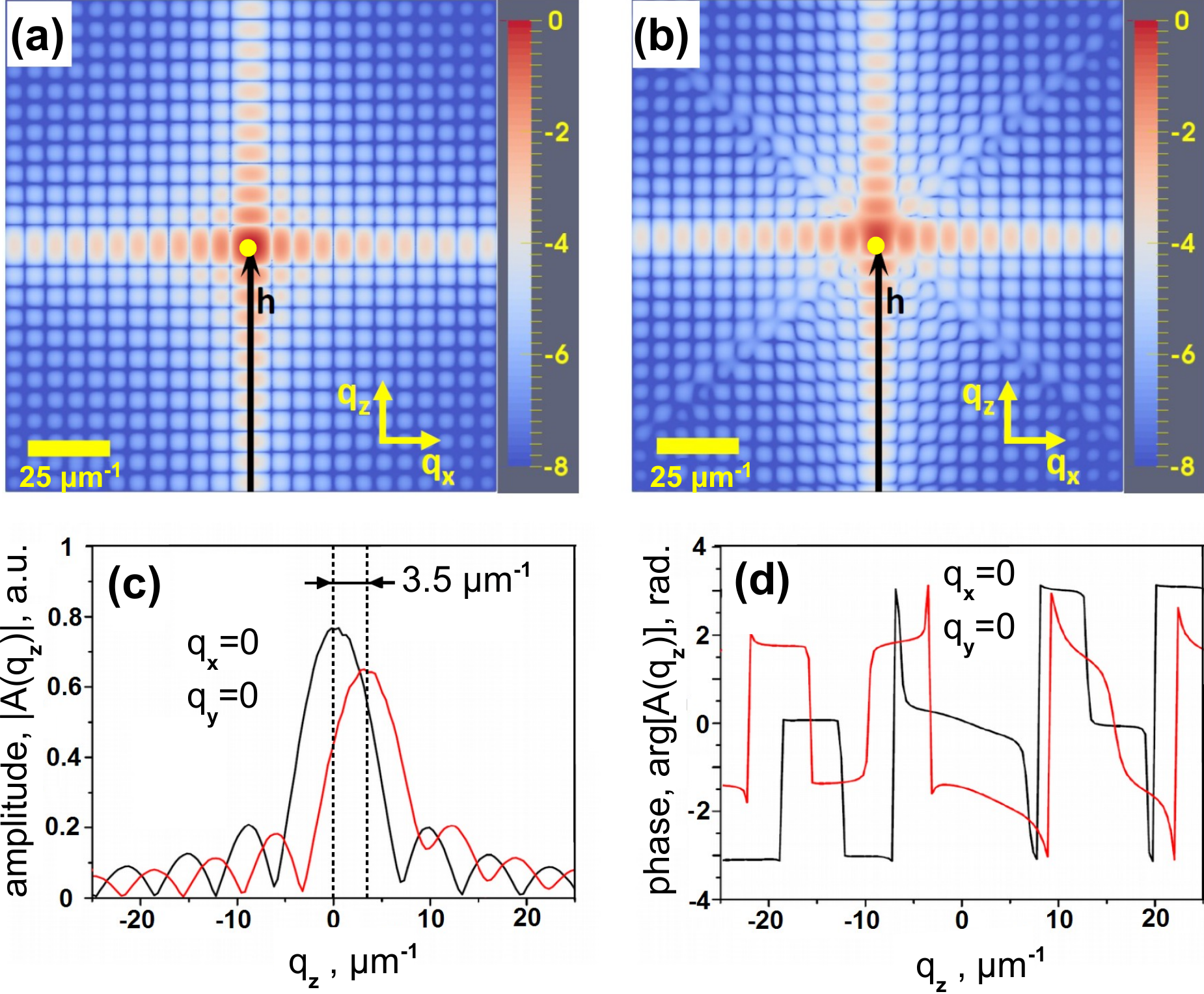}}
	\caption{
		(a, b) 2D distribution of the modulus of the scattered amplitude $\left|A(q_x, q_z)\right|$ taken through the central position ($q_y=0$) and simulated for a \SI{1}{\micro\meter} cubic Au crystal (shown in logarithmic scale).
        (a) Results of simulations performed in the frame of the kinematical theory,
		(b) results of the dynamical theory, obtained by a numerical solution of the Takagi-Taupin equations.
        Profiles of the modulus $\left|A(q_z)\right|$ and phase $\arg \left[ A(q_z)\right]$ along the $ q_z$-axis are shown in (c) and (d), respectively.
        Results of simulations performed in the frame of the kinematical theory (black lines) and dynamical theory (red lines).
        Note that only the central part of reciprocal space in the range from \SIrange{-25}{25}{\micro\meter^{-1}} is shown here.
}
	\label{fig:KinVsDynReci}
\newpage
\end{figure}

As a next step, we performed simulations for a \SI{1}{\micro\meter} Au crystal of a cubic shape.
Simulated 2D distribution of the modulus of the scattered amplitude $\left|A(q_x, q_z)\right|$ taken through the center of reciprocal space and obtained by using the kinematical and dynamical approaches is presented in Fig.~\ref{fig:KinVsDynReci} (a,b).
Two sets of crystal truncation rods perpendicular to the direction of the facets of the crystal as well as a regular structure of the square speckles due to coherent scattering on a cubic shape crystal are well seen in this Figure.
At the same time we see a significant difference between simulations performed with the kinematical and dynamical approaches.
The later ones show lower contrast and noticeable aberrations in the position and magnitude of the fringes.
We also observed an additional intensity in the form of a diagonal cross in the case of the dynamical theory simulations (see Fig.~\ref{fig:KinVsDynReci}(b)) that was also noticed in simulations performed in Ref.~\cite{punegov2016bragg}.


The difference in the position and intensity of the speckles is clearly seen in a linear scan of the amplitude $\left|A(q_z)\right|$
taken along the central rod (see Fig.~\ref{fig:KinVsDynReci}(c)).
A comparison of the kinematical (black line) and dynamical (red line) results show a displacement of the whole profile and particularly Bragg peak position in the positive direction of $q_z$-axis for the case of the dynamical theory simulations.
This result is well known in the dynamical theory~\cite{authier2001dynamical} and is due to refraction effect.
According to the dynamical theory the angular position of the maximum of the reflectivity curve is shifted from the exact Bragg position to
positive values by
\begin{equation}
\theta_{ref} = \mp\frac{\chi_{0r}(1\pm\beta)}{2\beta\sin2\theta_B}  \ ,
\label{eq:DarCurveDispl}
\end{equation}
where parameter $\beta=\gamma_0/|\gamma_h|$ for Bragg and $\beta=\gamma_0/\gamma_h$ for Laue geometries and the upper sign corresponds to Bragg diffraction and the lower one to Laue diffraction.
For Au(004), \SI{8}{\kilo\electronvolt} and symmetric Bragg geometry ($\gamma_0=|\gamma_h|$) the equation~\eqref{eq:DarCurveDispl} provides \SI{19.6}{\arcsecond} angular shift which is equivalent to $\SI{5.0}{\micro\meter}^{-1}$ of the positional displacement of the Bragg peak in reciprocal space along $q_z$ axis (compare with the similar results obtained in Ref.~\cite{punegov2016bragg}).
At the same time, in symmetric Laue geometry ($\gamma_0=\gamma_h$) no positional shift of the reflectivity curve is observed.
In the considered case of a cubic crystal the diffraction geometry is represented by a mixture of symmetric Bragg and Laue cases, therefore, refraction effects characteristic for these two geometries are superimposed.
That is revealed in a smearing of the central speckle in the direction of $q_z$-axis together with a positional shift of the maximum by \SI{3.5}{{\micro\meter}^{-1}} (see Fig.~\ref{fig:KinVsDynReci}(c)).
Although small angular displacement of the whole diffraction pattern due to refraction can be precisely determined in simulations it is rather challenging to consider it experimentally.
In most of experiments these effects are neglected and the maximum of the Bragg peak is assumed to be at an exact position of the reciprocal lattice node and is used as a reference position.

It is also well seen in Fig.~~\ref{fig:KinVsDynReci}(c) that due to the dynamical scattering contrast of the diffraction pattern is significantly reduced.
In experiment this might be erroneously attributed to lack of the transverse coherence and consequently partial coherence illumination~\cite{vartanyants2001partial, Williams_PRB_2007}, or vibrations of the sample stage.
While these effects may be compensated in reconstruction by the multimode decomposition~\cite{Clark_NatComm_2012, Thibault_Nature_2012} and attributed to the incoming field, however their physical origin is quite different and is due to the dynamical scattering effects.

In Fig.~\ref{fig:KinVsDynReci}(d) the corresponding $ q_z$-profiles of the phase distributions for the kinematical (black line) and dynamical (red line) calculations are presented.
Similar to the amplitude profiles shown in Fig.~\ref{fig:KinVsDynReci}(c) a comparison between the kinematical and dynamical results shows a positional displacement of the phase profile in the positive direction of $q_z$-axis in the case of the dynamical theory simulations.
In addition, the symmetry with respect to the positive and negative directions is broken and more complex structure of the profile is observed.

\begin{figure}
	\centerline{\includegraphics[angle=0, width=8.6cm, trim={0.9cm 8.5cm 1.0cm 1.0cm}, clip]{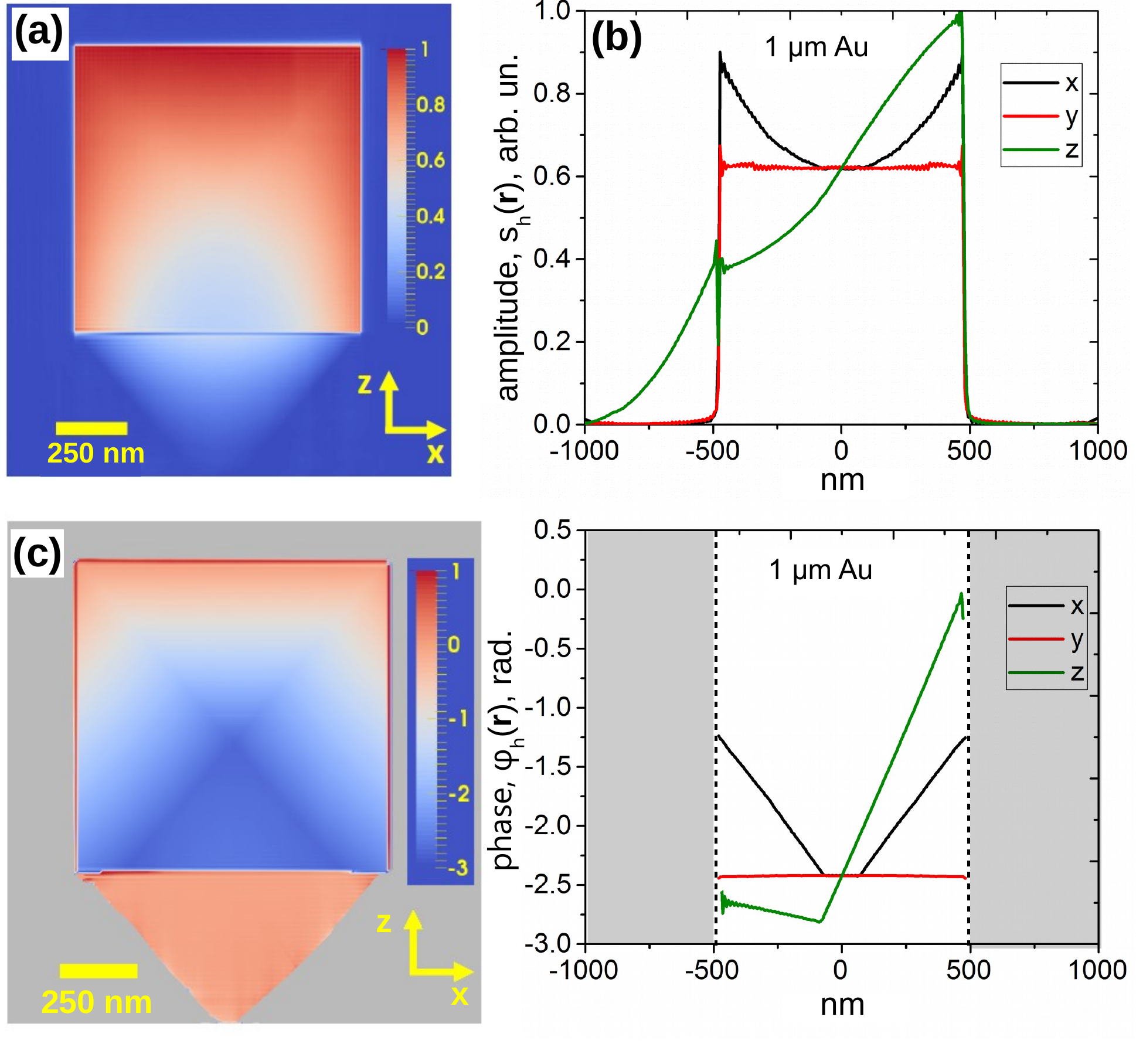}}
	\caption{
		Amplitude (a, b) and phase (c, d) of a complex crystalline function $S_h(\mathbf{r})$ obtained by inversion of the 3D reciprocal space dataset calculated for a crystal size of \SI{1}{\micro\meter}.
		(a, c) The $xz$-slices at $y=\num{0}$ (center of the crystal).
		(b, d) The line profiles through the center of the crystal and along the $x$, $y$, and $z$-axes.
		Gray area in (c, d) outlines a region outside the crystal, where the phase is undefined.
	}
	\label{fig:Cube1um}
\newpage
\end{figure}

The most intriguing were results of inversion performed for a \SI{1}{\micro\meter} size crystal.
In contrast to the previous case of a small crystal, results of inversion for a \SI{1}{\micro\meter} crystal (see Fig.~\ref{fig:Cube1um}) clearly show visible artifacts in the crystalline amplitude $s_h(\mathbf{r})$ (a,b) and phase $\varphi_h(\mathbf{r})$ (c,d) distribution in real space.
One strong effect, well visible in Fig.~\ref{fig:Cube1um}(a,b), is depletion of the crystalline amplitude towards the bottom of the crystal.
This is an expected effect of the dynamical theory.
Due to coupling of the incoming and diffracted waves at Bragg conditions the wave field is not propagating inside the crystal.
We observed that in our case the values of the amplitude dropped by more than 50\% (see Fig.~\ref{fig:Cube1um}(b)) instead of being uniform and constant on the level of one as in the case of a small crystal (see Fig.~\ref{fig:Cube100nm}(b)).
Another unexpected effect was appearance of an additional intensity, which extends below the bottom of the crystal (see Fig.~\ref{fig:Cube1um}(a,b)).
In our simulations for larger crystals (not shown) we observed that this artifact becomes stronger with the increase of the ratio of the crystal size to extinction length.
If the crystal shape is unknown before CXDI experiment such dynamical effects can result in a wrong reconstruction of the crystal shape as well as the values of the crystalline amplitude function.

Importantly, our simulations have revealed that the phase profile inside a crystal has a complicated distribution (see Fig.~\ref{fig:Cube1um}(c,d)).
We want to remind that initially we were considering Au crystals without any deformation.
At the same time, we obtained strong variations in the phase of the inverted crystalline function that should not be interpreted as originating from the crystal lattice deformation.
We will see in the following that some features of this phase distribution can be compensated by taking into account refraction effects.
Without such corrections the values of the strain field obtained from the CXDI reconstruction could be significantly different from the ones in the sample under investigation and in this way could bring to a wrong interpretation of the results in the Bragg CXDI experiment.

\section{ Treatment of refraction and absorption } 
\label{sec:corr}

Here we will analyse how effects of refraction and absorption could be taken into account.
We will perform analysis in the semi-kinematical approximation, when coupling between the incident and diffracted waves could be neglected but refraction and absorption effects will be taken into account specifically (see also Ref.~\cite{gorobtsov2016phase}).
We want to point out here that in conventional kinematical theory the incident and diffracted waves have no attenuation due to absorption and refraction effects are also neglected.

\begin{figure}
	\centerline{\includegraphics[angle=0, width=8.6cm, trim={1.0cm 8.5cm 1.0cm 1.0cm}, clip]{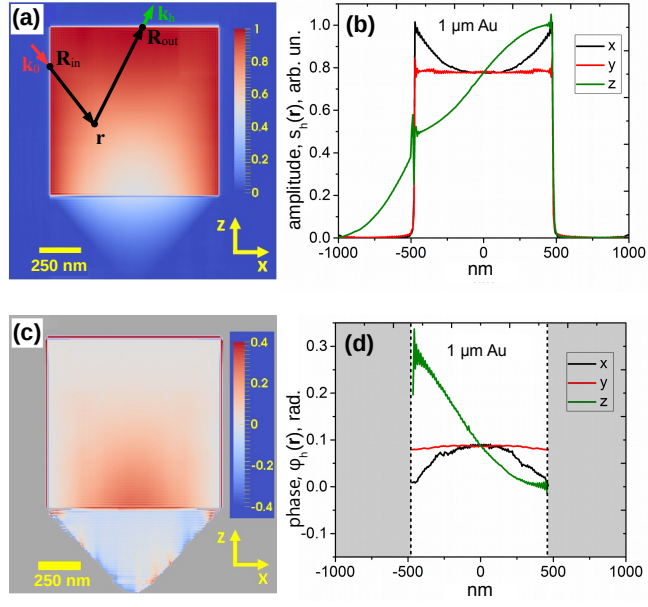}}
	\caption{
        Results of simulations for a \SI{1}{\micro\meter} Au crystalline particle presented in Fig.~\ref{fig:Cube1um} after applying  correction by the function $ f_c(\mathbf r) $ given by equation~\eqref{eq:CorrFunc}.
        The amplitude corrected for absorption (a, b) and the phase corrected for refraction (c, d) are represented for the $xz$-slice in (a, c) and by the line profiles along the $x$, $y$ and $z$-axes in (b, d).
        The sketch in (a) illustrates the total optical path $|\mathbf R_{in}-\mathbf r| + |\mathbf R_{out}-\mathbf r|$ calculated for a given point $\mathbf r$.
	    Gray area in (c, d) outlines region outside the crystal, where the phase is undefined.
	}
	\label{fig:Cube1umAftCorr}
\newpage
\end{figure}

To take all this into account, we will consider Takagi-Taupin equations \eqref{eq:TTMod} in which the coupling term in the first equation, proportional to $\chi_{\bar h}$, is eliminated that leads to the following system of equations
%
\begin{equation}
\begin{split}
\frac{\partial E_{0}(\mathbf r)}{\partial s_0}  =  \left(\frac{i\pi}{\lambda}\right) \chi_0 E_{0}(\mathbf r) \ , \\
\frac{\partial E_{h}(\mathbf r)}{\partial s_h}  =  \left(\frac{i\pi}{\lambda}\right) [\chi_0 E_{h}(\mathbf r) + \chi_h e^{i\Delta \mathbf q\cdot \mathbf r-i\mathbf h\cdot\mathbf u(\mathbf r)} E_{0}(\mathbf r)] \ .
\end{split}
\label{eq:TTMod_semkin}
\end{equation}
%
The first equation can be easily solved as
\begin{equation}
E_0(\mathbf{r}) = E_0^{in} \exp\left[i\frac{\chi_0}{2} \mathbf{k_0}(\mathbf{r}-\mathbf{R}_{in})\right] \ ,
\label{eq:Prop_wave}
\end{equation}
where $E_0^{in}$ is the incoming wavefield that will be put to unity in the following and $\mathbf{R}_{in} = \mathbf{R}_{in}(\mathbf{r})$ is a radius vector of the point, where the incoming beam enters the crystal for the considered element of volume (see sketch in Fig.~\ref{fig:Cube1umAftCorr}(a)).
Substituting this result in the second equation of the system of equations~\eqref{eq:TTMod_semkin} we obtain
\begin{equation}
\frac{\partial E_{h}(\mathbf r)}{\partial s_h} = \left(\frac{i\pi}{\lambda}\right) [\chi_0 E_{h}(\mathbf r) + \chi_h e^{i\Delta \mathbf q\cdot \mathbf r-i\mathbf h\cdot\mathbf u(\mathbf r)} e^{i\frac{\chi_0}{2} \mathbf{k_0}(\mathbf{r}-\mathbf{R}_{in})}] \ .
\label{eq:eq:TTMod_semkin_2nd}
\end{equation}
This equation for the diffracted wave can be solved by the following substitution
\begin{equation}
E_h(\mathbf{r}) =  E'_h(\mathbf{r}) e^{i\frac{\chi_0}{2} \mathbf{k_0}\mathbf{r}} \ ,
\label{eq:Prop_wave}
\end{equation}
which leads finally to the following analytical expression for the exit surface wave
\begin{equation}
E_h^{ESW}(\mathbf{R}_{out}) = \left(\frac{i\pi}{\lambda}\right)\chi_h \int S_h(\mathbf r) f_c(\mathbf{r})
e^{i \Delta \mathbf q \cdot \mathbf r} ds_h \ ,
\label{eq:ExitWaveDyn}
\end{equation}
where as before $S_h(\mathbf r)$ is a complex crystalline function and $ \mathbf{R}_{out}=\mathbf{R}_{out}(\mathbf r) $ is a radius vector of the position on a crystal surface where the diffracted beam exits the crystal for the considered element of volume (see sketch in Fig.~\ref{fig:Cube1umAftCorr}(a)).
We also introduced here a correction function
\begin{equation}
f_c(\mathbf{r}) = \left|f_c(\mathbf{r})\right| e^{i\varphi_c(\mathbf{r})}= \exp \left[i\frac{\chi_0}{2}\mathbf k_0 \cdot (\mathbf r-\mathbf R_{in})+i\frac{\chi_0}{2}\mathbf k_h \cdot (\mathbf R_{out}-\mathbf r)\right] \
\label{eq:CorrFunc}
\end{equation}
with its modulus due to absorption
\begin{equation}
\left|f_c(\mathbf{r})\right| = \exp \left[-\frac{\chi_{0i}}{2}\mathbf k_0 \cdot (\mathbf r-\mathbf R_{in}) - \frac{\chi_{0i}}{2}\mathbf k_h \cdot (\mathbf R_{out}-\mathbf r)\right]
\end{equation}
and the phase due to refraction
\begin{equation}
\varphi_c(\mathbf{r}) = \frac{\chi_{0r}}{2}\mathbf k_0 \cdot (\mathbf r-\mathbf R_{in}) + \frac{\chi_{0r}}{2}\mathbf k_h \cdot (\mathbf R_{out}-\mathbf r) \ .
\end{equation}
%
We want to point out again that in this treatment dynamical scattering effects are completely neglected.
Purely kinematical scattering can be directly obtained from Eqs.~(\ref{eq:ExitWaveDyn}-\ref{eq:CorrFunc}) by putting $\chi_0 = 0$ and consequently assuming that $f_c(\mathbf{r}) \equiv 1$.

As it follows from equations~(\ref{eq:ExitWaveDyn}-\ref{eq:CorrFunc}) the inversion of the reciprocal space dataset to real space should result in a complex function, which is represented by
$S_h(\mathbf{r}) f_c(\mathbf{r})$.
Therefore, in principle, to determine correctly the shape and strain field in a crystalline particle the correction function~\eqref{eq:CorrFunc} should be applied after inversion from reciprocal space.
In this correction function an optical path along the incident
$\mathbf k_0 \cdot (\mathbf r-\mathbf R_{in})$ and diffracted
$\mathbf k_h \cdot (\mathbf R_{out}-\mathbf r)$ beams should be calculated for each position $\mathbf{r}$ in a crystal (see sketch in Fig.~\ref{fig:Cube1umAftCorr}(a)).
An estimate is performed for a fixed angular position neglecting small variations of the optical path while the rocking scan.
Since the crystal shape and directions of the vectors $\mathbf{k}_0$ and $\mathbf{k}_h$ are known, the correction function~(\ref{eq:CorrFunc}) can be evaluated numerically in most of the cases.


Results of such correction applied to the complex electron density distribution obtained for a \SI{1}{\micro\meter} Au crystalline particle are shown in Fig.~\ref{fig:Cube1umAftCorr}.
The correction was performed only for positions inside a cubic volume, leaving the exterior part below the cube unchanged.
Comparison of the amplitudes reveals no significant changes in the upper part of the crystal.
In the lower part, where effects of absorption due to extinction are stronger, the values of the amplitude are increased from \num{0.38} to \num{0.52} (compare Fig.~\ref{fig:Cube1um}(a,b) and Fig.~\ref{fig:Cube1umAftCorr}(a,b)).
The correction revealed also a noticeable bump on the $z$-profile of the amplitude (see Fig.~\ref{fig:Cube1umAftCorr}(b)), which was barely pronounced in Fig.~\ref{fig:Cube1um}(b).
Still, major artifacts in the amplitude distribution, such as the depletion of the amplitude of the crystalline function in the bottom part of the crystal remained almost unchanged.
By that we conclude that remaining artifacts in the amplitude are related to purely dynamical effects in scattering.

\begin{figure}
	\centerline{\includegraphics[angle=0, width=8.6cm, trim={0.0cm 8.0cm 0.0cm 0.0cm}, clip]{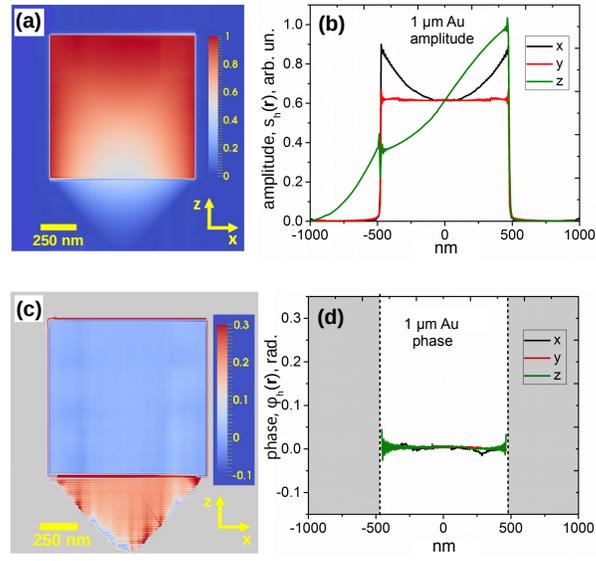}}
	\caption{
		Results of simulations for a \SI{1}{\micro\meter} Au crystalline particle calculated with an assumption of $ \chi_{hi} =
        \chi_{\bar hi}=\num{0}$, after applying correction by the function $ f_c(\mathbf r) $ given by equation~(\ref{eq:CorrFunc}).
		The amplitude (a, b) and phase (c, d) are represented for the $xz$-slice (at $y=0$) in (a, c) and by the line profiles along the $x$, $y$ and
        $z$-axes in (b, d).
		Gray area in (c, d) outlines region outside the crystal, where the phase is undefined.
	}	
	\label{fig:Cube1umHihReAftCorr}
\newpage
\end{figure}

At the same time, by applying correction function in the phase distribution we observed that a strong gradient of phase present in Fig.~\ref{fig:Cube1um}(c,d) is effectively removed (see Fig.~\ref{fig:Cube1umAftCorr}(c,d)).
Small residual aberrations in the range from \SIrange{0}{0.3}{\radian} are apparently connected to the dynamical effects~\cite{gorobtsov2016phase}.
More specifically, we determined that they can be attributed to the imaginary part of the Fourier components of the susceptibility $ \chi_h $ and $ \chi_{\bar h} $, which introduce a small phase shift when the wave is reflected by a crystalline plane.
To illustrate this we performed simulations for the same Au crystalline particle in which imaginary parts of the susceptibilities were eliminated from the Takagi-Taupin equations~(\ref{eq:TTMod}) by setting $\chi_{hi} = \chi_{\bar h i} = \num{0}$.
Results of these simulations after inversion to real space and applying correction by the function $f_c(\mathbf{r})$ (\ref{eq:CorrFunc}) are shown in Fig.~\ref{fig:Cube1umHihReAftCorr}.
As we can see from this figure all residual artifacts in the phase distribution were completely removed which approves our conclusion about the origin of these features.


To be sure that our correction function takes into account entire contribution of refraction and absorption we performed complementary simulations (not shown) where the corresponding terms were completely eliminated by setting $ \chi_0=\num{0} $ in Takagi-Taupin equations~(\ref{eq:TTMod}).
A comparison with the simulations performed by the fully dynamical case and corrections applied by the function $f_c(\mathbf{r})$ \eqref{eq:CorrFunc}, presented in Fig.~\ref{fig:Cube1umAftCorr}, showed that both results entirely coincide with each other.
This can be explained by a suggestion that contributions due to the dynamical scattering effects are completely decoupled from contributions originating from refraction and absorption.

\section{Simulations for a $Pb$ particle of a hemispherical shape}

Results of simulations for a perfect cubic Au crystalline particle have shown that the dynamical diffraction can lead to an appearance of artifacts in the real space reconstruction.
In order to estimate the contribution of the dynamical effects for a practical case, we considered experimental parameters described in Refs~\cite{pfeifer2006three, harder2007orientation}.
In that experiment 3D reconstruction of the Bragg CXDI data was used to characterize the strain distribution in a lead nanocrystal of a hemispherical shape of \SI{0.75}{\micro\meter} in diameter.
The crystal was coherently illuminated by a monochromatic x-ray beam of $\SI{1.38}{\angstrom}$ wavelength and Pb (111) reflection was selected.
In these experimental conditions the Bragg angle was \SI{13.97}{\degree} and values of the extinction length $L_{ex}$ were \SI{0.28}{\micro\meter} and \SI{1.14}{\micro\meter} for the Bragg and Laue geometries, respectively.

\begin{figure}
	\centerline{\includegraphics[angle=0, width=8.6cm, trim={1.2cm 16.2cm 1.5cm 1.2cm}, clip]{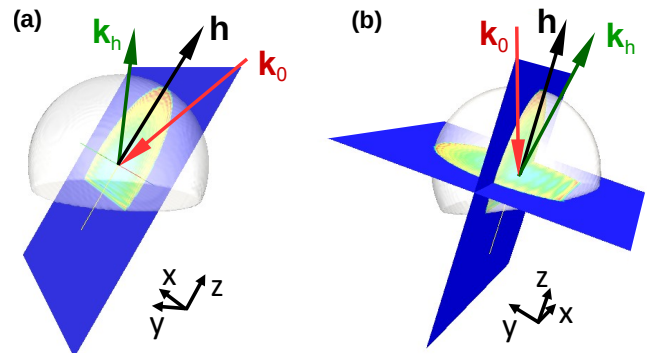}}
	\caption{
		Diffraction geometry used in simulations for a Pb crystalline nanoparticle of a hemispherical shape.
		Shape function is modeled by a sphere of \SI{0.75}{\micro\meter} in diameter truncated from one side by \num{01/3} of the diameter.
		The cutting plane is tilted by \SI{27}{\degree} with respect to (111) crystallographic plane.
        Two different perspectives are shown in (a) and (b).
		Blue planes outline the scattering plane ($xz$-slice) in (a) and (b) and diffraction plane ($xy$-slice) in (b).
	}
	\label{fig:SemiSphereSketch}
\newpage
\end{figure}

For our simulations we considered a shape function represented by a sphere truncated from one side by \num{1/3} of its diameter, as the closest model.
Following the description of the experiment, we oriented the truncation plane to form an angle of \SI{27}{\degree} with respect to the (111) crystallographic plane.
The diffraction geometry from two perspective views is schematically shown in Fig.~\ref{fig:SemiSphereSketch}.
Note orientation of the coordinate axes: similar to the case of simulations for a cubic Au particle the $x$- and $z$-axes lie in the scattering plane and the $y$-axis is orthogonal to them.
According to the chosen geometry (see Fig.~\ref{fig:SemiSphereSketch}) the cut along $ z $-axis is not symmetric with respect to the center but covers the range from $ -d/2+d/3=$\SI{-125}{\nano\meter} to $ d/2=$\SI{375}{\nano\meter}.

\begin{figure}
	\centerline{\includegraphics[angle=0, width=8.6cm, trim={1.0cm 10.5cm 1.0cm 1.0cm}, clip]{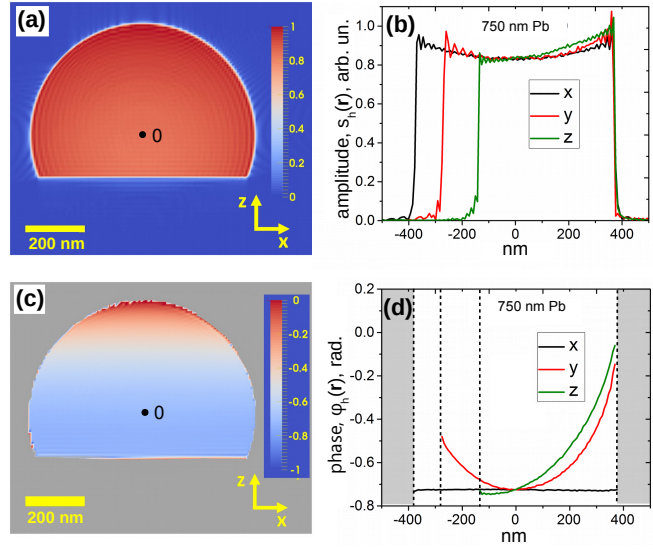}}
	\caption{
		Results of simulations for a hemispherical Pb crystal.
		3D distribution of the complex crystalline function was obtained by inversion of the scattered amplitudes calculated by the dynamical theory.
        The amplitude (a, b) and phase (c, d) are represented for the $xz$-slice in (a, c) and by the line profiles along the $x$, $y$ and
        $z$-axes in (b, d).
        Gray area in (c, d) outlines region outside the crystal, where the phase is undefined.
		}
	\label{fig:SemiSphere750nm}
\newpage
\end{figure}

Results of simulations performed by the dynamical theory are presented in Fig.~\ref{fig:SemiSphere750nm}.
A series of diffraction patterns were calculated in the angular range from \SI{-0.83}{\degree} to \SI{0.83}{\degree} with the
$3.3\cdot10^{-3}$ degrees angular increment.
They were merged into a 3D reciprocal space dataset, and then inverted to real space.
The amplitude distribution shown in Fig.~\ref{fig:SemiSphere750nm}(a, b) reveals slight depletion of the crystalline amplitude function $s_h(\mathbf{r})$ in the central part, which corresponds to attenuation of the incident and diffracted waves in the bulk of the crystal.
However, this artifact appears to be relatively small (about \SI{10}{\percent} of the average value) which supports applicability of the kinematical approach in this case.
At the same time, in the phase distribution (see Fig.~\ref{fig:SemiSphere750nm}(c,d)) a considerable phase gradient with the maximum deviation of the phase about \SI{0.7}{\radian} is observed.

\begin{figure}
	\centerline{\includegraphics[angle=0, width=8.6cm, trim={1.0cm 9.5cm 1.0cm 1.0cm}, clip]{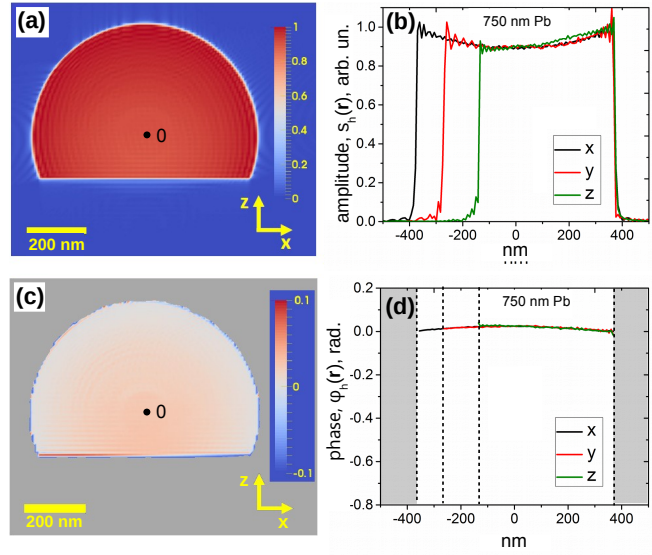}}
	\caption{
		Results of correction by the function $ f_c(\mathbf r) $ given by the equation~(\ref{eq:CorrFunc}) and applied to the results of
        simulations of a Pb semispherical particle presented in Fig.~\ref{fig:SemiSphere750nm}.
        The amplitude (a, b) and phase (c, d) are represented for the $xz$-slice in (a, c) and by the line profiles along the $x$, $y$ and
        $z$-axes in (b, d).
        Gray area in (c, d) outlines region outside the crystal, where the phase is undefined.
	}
	\label{fig:SemiSphere750nm_Cor}
\end{figure}

To reveal the origin of this phase gradient we applied the correction function $f_c(\mathbf{r})$~(\ref{eq:CorrFunc}) to the complex crystalline function $S_h(\mathbf{r})$ shown in Fig.~~\ref{fig:SemiSphere750nm}.
Results of this correction are presented in Fig.~\ref{fig:SemiSphere750nm_Cor}.
We do not observe any significant changes in the amplitude distribution (see Fig.~\ref{fig:SemiSphere750nm_Cor}(a, b)), as soon as contribution due to absorption is comparably small for this particle.
At the same time, correction due to refraction removed a major part of the gradient in the phase distribution (see Fig.~\ref{fig:SemiSphere750nm_Cor}(c, d)).
Leftover residual variations were on the level of \SI{0.03}{\radian} and can be neglected.
These values are much less than the values of the maximum phase deviation, which were estimated in Ref.~\cite{harder2007orientation} to be about \SI{1.15}{\radian} after correction for refraction effects.

Our results demonstrate, first, that corrections of the phase due to refraction are important even in the case of kinematical scattering and can not be neglected.
Second, our theoretical results demonstrate that an approach proposed in Ref.~~\cite{harder2007orientation} can be safely applied in the case when dynamical scattering effects are negligible and kinematical scattering approximation can be used.


\section{ Conclusions }
\label{sec:concl}

We present a general model based on a specific form of the Takagi-Taupin equations optimized for geometry of the Bragg CXDI measurement and with the aim to facilitate a numerical solution in a finite 3D crystal of an arbitrary shape in the presence of deformations.
As a result, the complex amplitude distributions of the transmitted and diffracted waves on the exit surface are calculated.
Propagation to the far-field provides the amplitude and phase distributions of the diffraction pattern that corresponds to a specific cross-section in reciprocal space.
By performing a series of such calculations for different values of rotation angle a full 3D reciprocal space dataset in the vicinity of the corresponding reciprocal lattice node can be constructed.
The complex crystalline function of the object in real space is obtained by the inverse Fourier transform.

Using this model we performed simulations of the dynamical diffraction on a perfect crystal of gold of a cubic shape of \SI{100}{\nano\meter} and \SI{1}{\micro\meter} in size.
For a small crystal results of our calculations were in full agreement with the kinematical theory.
In the simulations for the large crystal artifacts introduced by the dynamical scattering effects were observed in real as well as in reciprocal spaces.
We analyzed the contributions of different phenomena, such as refraction, absorption and cross-coupling between the diffracted and transmitted waves.
Based on the analytical derivations we developed an approach which corrects the results of reconstructions for the effects of refraction and absorption.
Such correction, applied to the results of the simulations, demonstrates a complete removal of corresponding contributions in the real space reconstruction.
The residual artifacts in the amplitude and phase distributions are attributed to the dynamical effects of scattering in the crystal.
Additional simulation for a practical case of a Bragg CXDI experiment with a hemispherical Pb particle of \SI{750}{\nano\meter} in size was also performed.
By applying the correction for refraction and absorption we demonstrate that remaining dynamical artifacts were small and did not affect results of the reconstruction.

We conclude that limitations of kinematical approach in the Bragg CXDI experiments depend on the relative values of the crystal size $d$ and extinction length $L_{ex}$.
We suggest the following critera.
If scattering conditions (crystal shape and orientation) are predominantly Bragg (Laue) than the size of the crystal $d$ should be compared with the corresponding Bragg (Laue) extinction length.
When the crystal size is smaller than the corresponding extinction length, dynamical effects should not affect reconstruction significantly.
However, even in this case effects of absorption and especially refraction should be specially analyzed.
If necessary, correction function should be applied to determine correct values of strain in the sample.
In other cases the dynamical theory should be applied.

Finally, we think that our findings will be of high importance for all groups working in the fast developing field of coherent scattering and imaging in Bragg scattering conditions.

\bigskip

\begin{acknowledgments}

Authors are grateful to E. Weckert for fruitful discussions and support of the project.
We acknowledge careful reading of the manuscript and helpful discussions with O. Gorobtsov and I. Sergeev.
This research was partially supported by the Virtual Institute VH-VI-403 of the Helmholtz Association.

\end{acknowledgments}

\appendix

\section{Derivation of modified Takagi-Taupin equations}
\label{app:AppT}

Here we describe derivation of the Takagi-Taupin equations in symmetric form.
The properties of the electric field vector $\mathbf E(\mathbf r)$ inside a crystal are described by the following wave propagation equation %
\begin{equation}
\Delta \mathbf E(\mathbf r)  -\si{grad}(\si{div}\mathbf E(\mathbf r))+ \frac{\omega^2}{c^2}[1+\chi (\mathbf r)]\mathbf E(\mathbf r) = 0 \ ,
\label{eq:WaveEq}
\end{equation}
where $\omega$ is the frequency of the wave field and $c$ is the speed of light.
In equation~\eqref{eq:WaveEq} $\chi (\mathbf r)$ is the susceptibility of the crystal.
We assume in the following the case of $ \sigma-$polarization, so the electric field will be further considered as a scalar field.

In the case of a perfect crystal the susceptibility $ \chi ( \mathbf r)$ is a periodic function with the period of the crystal lattice that can be expanded as a Fourier series
\begin{equation}
\chi^{(id)}(\mathbf{r}) = \sum _{h}\chi^{(id)}_{h} e^{i\mathbf{h}\cdot\mathbf{r}} \ ,
\label{eq:HiExpan}
\end{equation}
where $\mathbf{h}$ is the reciprocal lattice vector.
In equation~\eqref{eq:HiExpan} the summation is carried out over all reciprocal lattice vectors.
%
%
In the case of weak deformations, when relative displacements are small the susceptibility of the crystal $\chi(\mathbf{r})$ is defined from that of a perfect one according to the relation~\cite{takagi1962dynamical}, $\chi(\mathbf{r}) = \chi^{(id)}(\mathbf{r}-\mathbf{u}(\mathbf{r}))$. The Fourier components of the susceptibility in the weakly deformed crystal now depend on the coordinate $\mathbf{r}$ and can be defined as

%
\begin{equation}
\chi_h(\mathbf{r}) = \chi_h^{(id)} e^{-i\mathbf{h}\cdot\mathbf{u}(\mathbf{r})} \ .
\label{eq:HiHDeformed}
\end{equation}
%

The solution of equation~\eqref{eq:WaveEq} may be found in the form of an expansion analogous to Bloch waves (see equation~\eqref{eq:TwoWaveApp}) and leads to a well known form of the Takagi-Taupin equations (see for example~\cite{authier2001dynamical, vartaniants2001XSW}).

We will consider now that the orientation of the crystal satisfies the exact Bragg conditions.
In this case we have for the Takagi-Taupin equations
%
\begin{equation}
\begin{split}
\frac{\partial E_{0}(\mathbf{r})}{\partial s_{0}} = \frac{i\pi}{\lambda} [\chi _{0} E_{0}(\mathbf{r})+\chi _{\bar{h}}e^{i\mathbf h\cdot\mathbf u(\mathbf r)} E_h (\mathbf{r})] \ ,\\
\frac{\partial E_h(\mathbf{r})}{\partial s_h } = \frac{i\pi}{\lambda} [\chi_0 E_h (\mathbf{r})+\chi _{h} e^{-i\mathbf h\cdot\mathbf u(\mathbf r)} E_{0} (\mathbf{r})] \ .
\end{split}
\label{eq:TTBragg}
\end{equation}
%

We assume now that the crystal is rotated by an angle $\Delta \theta$, and denote reciprocal lattice vector at this new orientation as $\mathbf h' = \mathbf{h} + \Delta \mathbf{q}$.
Then, Fourier decomposition of the susceptibility $ \chi' ( \mathbf r)$ for this new angular position of a crystal may be written as
\begin{equation}
\chi'(\mathbf{r}) = \sum _{h} [\chi_{h} e^{i(\mathbf{h'}-\mathbf{h})\cdot\mathbf{r}}] e^{i\mathbf{h}\cdot\mathbf{r}} \ .
\label{eq:HiExpan2}
\end{equation}
Comparing this expression with the decomposition~(\ref{eq:HiExpan}) we conclude that in the equations~(\ref{eq:TTBragg}) the following  substitutions should be made
\begin{equation}
\begin{split}
\chi_h \to \chi_h e^{i(\mathbf{h'}-\mathbf{h})\cdot \mathbf{r}}  \to \chi_h e^{i \Delta \mathbf{q} \cdot \mathbf{r}} \\
\chi_{\bar h} \to \chi_{\bar h} e^{i(\mathbf{h}-\mathbf{h'})\cdot \mathbf{r}} \to \chi_{\bar h} e^{-i\Delta \mathbf{q} \cdot \mathbf{r}}\ .
\end{split}
\label{eq:HiHSubst}
\end{equation}
Following this approach, the Takagi-Taupin equations for this new crystal orientation can be finally written in the form~(\ref{eq:TTMod}).
The angular dependence in the set of equation~(\ref{eq:TTMod}) is represented by the phase exponent $ \exp(i\Delta \mathbf q\cdot \mathbf r)$ which leads to convenient and symmetric form of the Takagi-Taupin equations used in the simulations.


\section{Takagi-Taupin equations for the modified amplitudes}

In Takagi-Taupin equations~\eqref{eq:TTMod} two coupling terms with $\chi_{h,\bar h}$ are responsible for dynamical diffraction effects.
If we eliminated these terms from equations~(\ref{eq:TTMod}), they turn into a linear independent differential equations, which describe an independent transmission of the waves $E_{0,h}(\mathbf r)$ through a crystal without diffraction.
In this case, the analytical solution for each equation is represented by an exponential function $\exp(i\pi\chi_0s_{0,h}/\lambda)$.

Following this approach we substitute the amplitudes $E_{0,h}(\mathbf r)$ in the Takagi-Taupin equations~\eqref{eq:TTMod}  by the new ones defined as (compare with Ref.\cite{gorobtsov2016phase})
\begin{equation}
\begin{split}
E_{0, h}(\mathbf r)=E'_{0, h}(\mathbf r)e^{i\frac{\chi_0}{2}\mathbf k_{0, h} \cdot \mathbf r} \ ,
\end{split}
\label{eq:ModAmpl}
\end{equation}
which lead us to a new system of equations
\begin{equation}
\begin{split}
\frac{\partial E'_0(\mathbf r)}{\partial s_0} = \frac{i\pi}{\lambda}\chi_{\bar h} e^{-i\Delta \mathbf q'\cdot\mathbf r+i\mathbf h\cdot\mathbf u(\mathbf r)} E'_h(\mathbf r) \ ,\\
\frac{\partial E'_h(\mathbf r)}{\partial s_h} = \frac{i\pi}{\lambda} \chi_h e^{i\Delta \mathbf q'\cdot\mathbf r-i\mathbf h\cdot\mathbf u(\mathbf r)} E'_0(\mathbf r) \ ,
\end{split}
\label{eq:TTModComp}
\end{equation}
where the complex vector $\Delta \mathbf q'$ is defined  as
\begin{equation}
\Delta \mathbf q'=\Delta \mathbf q+\frac{\chi_0}{2}(\textbf k_0-\textbf k_h) \ .
\label{eq:DQ}
\end{equation}

This approach allows one to consider products in exponential factors as additives to the wave vectors and treat those as complex values with the directional properties given only by their real parts~\cite{authier2001dynamical}.
As such the refraction and absorption of both diffracted and transmitted waves are included in their definition by applying boundary conditions for these amplitudes on a crystal surface.
%
%
As it naturally follows from expressions~(\ref{eq:ModAmpl}), the modified amplitudes $E'_{0,h}(\mathbf r)$ differ from the amplitudes $E_{0,h}(\mathbf r)$ inside the material, but are the same in a vacuum.
Therefore, the boundary conditions for equations~(\ref{eq:TTModComp}) should be expressed as
\begin{equation}
E_{0,h}^{'vac}(\mathbf R_b) =  E_{0,h}^{'cryst}(\mathbf R_b)e^{i\frac{\chi_0}{2}\mathbf k_{0,h}\cdot\mathbf R_b} \ ,
\label{eq:BoundCondDQ}
\end{equation}
where $\mathbf R_b$ is the radius vector of a considered point at the crystal-vacuum boundary, $E_{0,h}^{'vac}(\mathbf r)$ and  $E_{0,h}^{'cryst}(\mathbf r)$ are the values of amplitudes in a vacuum and inside the crystal.
In fact, that is equivalent to the condition of continuity of the tangential component of the electric field at the interface.
Such approach allows to treat effects of refraction and absorption while propagating the wavefields in a crystal in a simple way.

\section{ Numerical matrix form of the Takagi-Taupin equations}
\label{app:num}

In the numerical integration method the complex amplitudes $E_{0,h}(\mathbf r)$ are represented by a discrete set of values over all integration grid and the Takagi-Taupin equations are transformed to a recurrence matrix form, similar to Ref.~\cite{gronkowski1991propagation}.
Relying upon the symmetry of equations between the transmitted and the diffracted amplitudes, we take the same elementary integration step $p$ for both directions.
For any smooth, slowly varying function $f(x, y)$ one can use the finite difference approximation to estimate the partial derivative from values at two neighboring points
\begin{equation}
\frac{\partial}{\partial x}f(x-\frac{p}{2}, y)=\frac{f(x, y)-f(x-p, y)}{p} \ .
\label{eq:CenDiffAppr}
\end{equation}
The value of the function in this middle point is given by a half sum
\begin{equation}
f(x-\frac{p}{2}, y)=\frac{f(x, y)}{2}+\frac{f(x-p, y)}{2} \ .
\label{eq:HalfSumApp}
\end{equation}
When these formulas are applied to the differential equations~(\ref{eq:TTMod}) those are transformed to the following set
\begin{equation}
\begin{split}
E_0(s_0,s_h)-E_0(s_0-p,s_h) = \frac{i\pi}{\num{2}\lambda} [\chi_0 E_0(s_0,s_h)+ \\ +\chi_0 E_0(s_0-p,s_h)+B E_h(s_0,s_h)+B E_h(s_0-p,s_h)] \ , \\
E_h(s_0,s_h)-E_h(s_0,s_h-p) = \frac{i\pi}{\num{2}\lambda} [\chi_0 E_h(s_0,s_h)+ \\ \chi_0 E_h(s_0,s_h-p)+D E_0(s_0,s_h)+D E_0(s_0,s_h-p)] \ ,
\end{split}
\label{eq:TTonIntNet}
\end{equation}
with substitutions
\begin{equation}
\begin{split}
B=\chi_{\bar h}\exp [-i(s_0-\frac{p}{2})\Delta \mathbf q\cdot\mathbf s_0+i s_h\Delta \mathbf q\cdot\mathbf s_h+ \\
+ i\mathbf h\cdot\mathbf u(s_0-\frac{p}{2},s_h)] \ , \\
D=\chi_h \exp [is_0\Delta \mathbf q\cdot\mathbf s_0-i(s_h-\frac{p}{2})\Delta \mathbf q\cdot\mathbf s_h - \\
- i\mathbf h\cdot\mathbf u(s_0,s_h-\frac{p}{2})] \ .
\end{split}
\label{eq:BD}
\end{equation}
All the considered points belong to the same scattering plane, therefore in further derivations we simply omit the $s_y$ coordinate in the aid of shortness.
The set of equations~(\ref{eq:TTonIntNet}) can be further reorganized to
\begin{equation}
\begin{split}
E_0(s_0,s_h)=\frac{A}{C}E_0(s_0-p,s_h)+\frac{B}{C}E_h(s_0,s_h)+ \\
+ \frac{B}{C}E_h(s_0-p,s_h) \ ,\\
E_h(s_0,s_h)=\frac{A}{C}E_h(s_0,s_h-p)+\frac{D}{C}E_0(s_0,s_h)+ \\
+ \frac{D}{C}E_0(s_0,s_h-p) \ .
\end{split}
\label{eq:TTonIntNet_Next}
\end{equation}
Here two additional substitutions were made
\begin{equation}
\begin{split}
A=\frac{\num{2}\lambda}{i\pi p}+\chi_0 \ , \\
C=\frac{\num{2}\lambda}{i\pi p}-\chi_0 \ ,
\end{split}
\label{eq:AC}
\end{equation}
and it was assumed that $C\neq0$, which is evidently true for any real positive $ p $ as far as $\chi_{0r}\neq 0$.
Solving this system with respect to $E_0(s_0,s_h)$ and $E_h(s_0,s_h)$ we obtain
\begin{equation}
\begin{split}
E_0(s_0,s_h)[1-\frac{BD}{C^{2}}] = \frac{A}{C}E_0(s_0-p,s_h)+ \\ +\frac{B}{C}E_h(s_0-p,s_h)+\frac{BD}{C^2}E_0(s_0,s_h-p)+ \\ +\frac{BA}{C^2}E_h(s_0,s_h-p) \ , \\
E_h(s_0,s_h)[1-\frac{BD}{C^2}] = \frac{AD}{C^2}E_0(s_0-p,s_h)+ \\ \frac{BD}{C^2}E_h(s_0-p,s_h)+\frac{D}{C}E_0(s_0,s_h-p)+ \\ +\frac{A}{C}E_h(s_0,s_h-p) \ .
\end{split}
\label{eq:TT_recurrent}
\end{equation}
These relations can be also written in the matrix form
\begin{equation}
\begin{pmatrix}E_0(s_0,s_h) \\ E_h(s_0,s_h)\end{pmatrix} = M \begin{pmatrix}E_0(s_0-p,s_h) \\ E_h(s_0-p,s_h) \\ E_0(s_0,s_h-p) \\ E_h(s_0,s_h-p)\end{pmatrix} \ ,
\label{eq:TT_matrix}
\end{equation}
where coefficients of matrix $ M $ are expressed as
\begin{equation}
M = \frac{\num{1}}{C^2-BD}
\begin{pmatrix}AC\quad BC\quad BD\quad BA \\ AD\quad BD\quad DC\quad AC\end{pmatrix} \ .
\label{eq:TT_matrixM}
\end{equation}

\section{ Kinematical limit of the Takagi-Taupin equations}
\label{app:AnSol}

\begin{figure}[p]
	\centerline{\includegraphics[angle=0, width=8.6cm, trim={0.0cm 6.7cm 0.0cm 0.5cm}, clip]{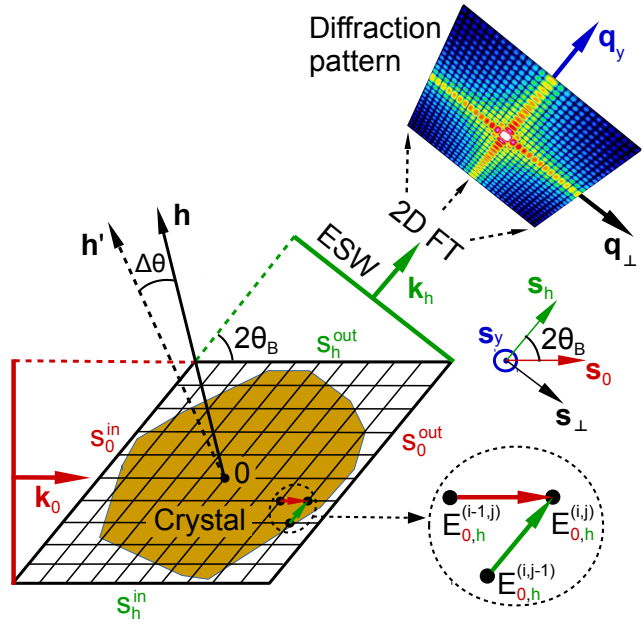}}
	\caption{
		Schematics of the numerical model used for simulations of 2D Bragg CXDI diffraction from a finite crystal.
		Calculations are performed for each value of the angular deviation $\Delta \theta$.
        Crystal is rotated around the axis going through the crystal center (denoted as 0) that is perpendicular to the scattering plane defined by the vectors $\mathbf{k}_0$ and $\mathbf{k}_h$.
        See text for the details of simulations.
    The inset on the right shows recurrent relations for a single node of the grid.
	}
	\label{fig:Scheme_NumInt}
\newpage
\end{figure}

Here we derive an analytical solution of the Takagi-Taupin equations for the purely kinematical case.
In equations~\eqref{eq:TTMod} we neglect coupling between the transmitted and diffracted components of the wave field, which is described by the term $\chi_{\bar{h}}$, we also neglect effects of refraction and absorption which are described by the term $\chi_{0}$.
This leads to the following form of the Takagi-Taupin equations~\eqref{eq:TTMod}
\begin{equation}
\begin{split}
\frac{\partial E_0(\mathbf r)}{\partial s_0} = 0  \ , \\
\frac{\partial E_h(\mathbf r)}{\partial s_h} = \left(\frac{i\pi}{\lambda}\right) \chi_h e^{i\Delta \mathbf q\cdot \mathbf r-i\mathbf h\cdot\mathbf u(\mathbf r)} E_0(\mathbf r) \ .
\end{split}
\label{eq:EdKin}
\end{equation}
The first equation can be easily solved as $E_0(\mathbf r) = 1$, where we assumed that the amplitude of the incoming beam is equal to unity.
The amplitude of the diffracted wave field at the exit surface of the crystal can be obtained from the second equation in~\eqref{eq:EdKin}
%
\begin{equation}
E_h^{ESW}(s_{\bot },s_y, \Delta \mathbf q) = \left(\frac{i\pi}{\lambda}\right)\chi_h \int S_h(\mathbf r) e^{i \Delta \mathbf q\cdot\mathbf r} ds_h \ ,
\label{eq:ExitWaveKin}
\end{equation}
where the following representation of the position vector $\mathbf r = s_{\bot}\mathbf s_{\bot}+s_y\mathbf s_y+s_h\mathbf s_h$ is used (see Fig.~\ref{fig:Scheme_NumInt}) and $S_h(\mathbf r)$ is a complex crystalline function.
As it was discussed before the far-field diffraction pattern can be obtained by the 2D Fourier transform of the exit surface wave
\begin{equation}
\begin{split}
A(q_{\bot},q_y, \Delta \mathbf q) = \\
= \iint E_h^{ESW}(s_{\bot},s_y, \Delta \mathbf q) e^{-iq_{\bot}s_{\bot}-iq_y s_y}ds_{\bot}ds_y \ ,
\end{split}
\label{eq:FT_EW_Kin}
\end{equation}
where $q_{\bot}, q_y$ are the reciprocal space coordinates.
By substituting an expression~(\ref{eq:ExitWaveKin}) in equation~(\ref{eq:FT_EW_Kin}) we obtain
\begin{equation}
\begin{split}
A(q_{\bot }, q_y, \Delta \mathbf q) = \\
=\left(\frac{i\pi}{\lambda}\right) \chi_h \iiint S_h(\mathbf r) e^{i\Delta \mathbf q\cdot\mathbf r-iq_{\bot } s_{\bot } -iq_{y} s_{y} } ds_{\bot } ds_y ds_h \ .
\end{split}
\label{eq:DiffPattKin}
\end{equation}
Since $ ds_{\bot } ds_y ds_h =d \mathbf r$, the integral on the right side of equation~\eqref{eq:DiffPattKin} is the 3D Fourier transform of a complex crystalline function
$S_h(\mathbf{r}) = s_h(\mathbf r)\exp(-i\mathbf h\cdot\mathbf u(\mathbf r))$.
Comparison of equation~\eqref{eq:DiffPattKin} with equation~\eqref{eq:ScattEq} shows that they completely coincide, which gives a confidence that our approximations to the Takagi-Taupin equations indeed correspond to the kinematical diffraction case.


\bibliography{./references}

\eject


\end{document}